\documentclass[sigconf]{acmart}
\usepackage{booktabs} 
\setcopyright{rightsretained}
\acmDOI{10.475/123_4}

\acmISBN{123-4567-24-567/08/06}

\renewcommand\footnotetextcopyrightpermission[1]{} 
\usepackage{url}
\usepackage{graphicx}
\usepackage{cuted}
\usepackage{tabularx}
\usepackage{amsmath, amssymb, dsfont, cuted}
\usepackage{listings,multicol, multirow}
\usepackage{array}
\usepackage{color}
\usepackage{diagbox}
\newcolumntype{L}[1]{>{\raggedright\let\newline\\\arraybackslash\hspace{0pt}}m{#1}}
\newcolumntype{C}[1]{>{\centering\let\newline\\\arraybackslash\hspace{0pt}}m{#1}}
\newcolumntype{R}[1]{>{\raggedleft\let\newline\\\arraybackslash\hspace{0pt}}m{#1}}

\newtheorem{remark}{Remark}

\newcommand{\vF}{v}
\newcommand{\vFdes}{v^{\rm des}}
\newcommand{\vFmin}{v^{\rm min}}
\newcommand{\vFmax}{v^{\rm max}}

\newcommand{\headway}{h}
\newcommand{\hmin}{h^{\rm min}}

\newcommand{\vL}{v_L}
\newcommand{\vLmin}{v_L^{\rm min}}
\newcommand{\vLmax}{v_L^{\rm max}}

\newcommand{\aL}{a_L}
\newcommand{\aLmin}{a_L^{\rm min}}
\newcommand{\aLmax}{a_L^{\rm max}}

\newcommand{\Fw}{F_w}
\newcommand{\Fwmin}{F_{w,c}^{\rm \min}}
\newcommand{\Fwmax}{F_{w,c}^{\rm max}}

\newcommand{\Fwminphys}{F_{w,p}^{\rm \min}}
\newcommand{\Fwmaxphys}{F_{w,p}^{\rm max}}

\newcommand{\wdes}{\omega^{\rm des}}
\newcommand{\wmin}{\omega^{\rm min}}

\newcommand{\ldev}{y}
\newcommand{\vLateral}{\nu}
\newcommand{\adev}{\Delta\Psi}
\newcommand{\yaw}{r}

\newcommand{\vN}{v_N}
\newcommand{\xLK}{x_{LK}}

\makeatletter
\def\munderbar#1{\underline{\sbox\tw@{$#1$}\dp\tw@\z@\box\tw@}}
\def\moverbar#1{\overline{\sbox\tw@{$#1$}\dp\tw@\z@\box\tw@}}
\makeatother

\newcommand{\diag}{\text{diag}}
\newcommand{\comma}{\texttt{Comma AI}}
\newcommand{\commastar}{\texttt{Comma AI*}}
\newcommand{\staliro}{\texttt{S-TaLiRo}}

\def\beq{\begin{equation}}
\def\eeq{\end{equation}}

\DeclareMathOperator*{\argmax}{\arg\!\max}

\begin{document}

\thispagestyle{empty}

\title{Using control synthesis to generate corner cases: A case study on autonomous driving}

\author{Glen Chou$^1$, Yunus E. Sahin$^1$, Liren Yang$^1$, Kwesi J. Rutledge$^1$, Petter Nilsson$^2$, Necmiye Ozay$^1$\\
$^1$ University of Michigan, Ann Arbor\\
$^2$ California Institute of Technology, Pasadena}
\thanks{The first three authors contributed equally.}

\begin{abstract}

This paper employs correct-by-construction control synthesis, in particular controlled invariant set computations, for falsification. Our hypothesis is that if it is possible
to compute a ``large enough" controlled invariant set either for the actual system model
or some simplification of the system model, interesting
corner cases for other control designs can be generated by sampling initial conditions from the boundary of
this controlled invariant set. Moreover, if falsifying trajectories for a given control design
can be found through such sampling, then the controlled invariant set can be used as 
a supervisor to ensure safe operation of the control design under consideration. In addition to interesting initial conditions, which are mostly related to safety violations in transients, we use solutions from a dual game, a reachability game for the safety specification, to find falsifying inputs. We also propose optimization-based heuristics for input generation for cases when the state is outside the winning set of the dual game.
To demonstrate the proposed ideas, we consider case studies from 
basic autonomous driving functionality, in particular,
adaptive cruise control and lane keeping. We show how the proposed technique can be used to find interesting
falsifying trajectories for classical control designs like proportional controllers, proportional integral controllers and 
model predictive controllers, as well as 
an open source real-world autonomous driving package.

\end{abstract}


\maketitle

\section{Introduction} 
\label{sec:introduction}

Formal verification, the process of algorithmically generating correctness certificates for a design, and falsification, the process of algorithmically finding trajectories and inputs that lead to a violation of specifications are important steps before a safety-critical control system can be deployed \cite{fainekos2012verification,annpureddy2011s,sankaranarayanan2012falsification}. An alternative to these approaches, when a control design is not available but a plant model and specifications are available, is to synthesize a controller that, by construction, guarantees that the specifications are satisfied by the closed-loop system \cite{ozay2017guest}. The key insight of this paper is to combine ideas from falsification and control synthesis to evaluate control designs for safety.

Consider the problem of evaluating a control design for an autonomous vehicle for safety. What would be a meaningful specification to run a falsification engine against in this case? The hard safety constraint -- ``do not crash!" -- is easy to specify but can be trivially falsified. For instance, if a lead car, with very low speed, cuts in front of the autonomous car traveling with a relatively high speed, a crash is unavoidable. To get ``interesting'' corner cases, one might constrain the distance at which the lead car cuts in or the speed the lead car is traveling at when it cuts in. But can we systematically generate such constraints/assumptions? If a falsifying trajectory is found, can we say anything about existence of a controller that would be able to steer the vehicle to safety, or is safety simply an impossible task in this situation?  

Motivated by these questions, in this paper we propose to use controlled invariant sets \cite{blanchini1999survey} to generate interesting corner cases for falsification. By an interesting corner case, we mean initial conditions from which ensuring safety is hard but not necessarily impossible. We restrict our attention to piecewise affine control systems subject to external disturbances (e.g., behavior of the other cars on the road, road profile) and safety constraints given as unions of polyhedra. We propose a scheme to sample initial conditions from the boundary of the invariant set. We also consider the problem of searching for falsifying disturbances (in addition to initial conditions). To this effect, we compute the winning set of a dual game, where control inputs are treated as disturbances and disturbances are treated as control, and where the goal is to reach the unsafe set. The dual strategy obtained by solving the dual game can be used to generate falsifying inputs when the state is within the winning set of the dual game. Greedy heuristics that aim to push the states to the dual game winning set are also proposed.

As an additional advantage, in case a control design is found unsafe using the proposed method, we can supervise this unsafe controller with the controlled invariant set in order to guarantee safety while still using the unsafe controller, which may have favorable performance related properties \cite{nilsson2016correct}. This supervision idea is similar to the simplex architecture \cite{bak2011sandboxing,seto1998dynamic}, where a performance controller is used together with a simpler controller that has a certified safety envelope and that overwrites the performance controller only when its actions risk safety.

We demonstrate the proposed approach using two autonomous driving functions: adaptive cruise control and lane keeping. Adaptive cruise control aims to regulate the longitudinal dynamics of a vehicle either to a desired speed or a desired headway to a lead vehicle. Lane keeping controls the lateral dynamics of a vehicle to track the center line of the lane.  We present safety specifications for both functions. We then apply the proposed approach to a set of controllers, including \comma{} software, an open source autonomous driving package, to reveal potential corner cases leading to specification violation.

\section{Main Ingredients}

\begin{figure}
  \begin{center}
    \includegraphics[width=\columnwidth]{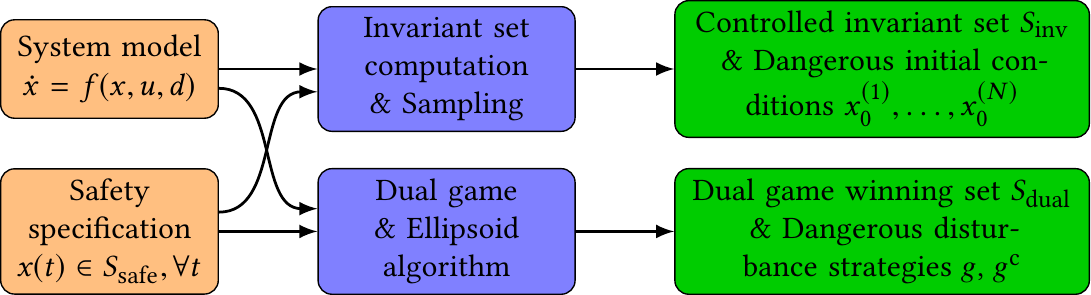}
  \end{center}
  \caption{Main workflow. Given a system model and a safety specification we synthesize a controlled invariant set contained inside the safe set and a winning set for the dual game. Based on these two objects we extract ``interesting" initial conditions and disturbance strategies that are used to evaluate the safety of arbitrary (black-box) controllers.}
  \label{fig:workflow}
\end{figure}

\begin{figure}[htb]
  \begin{center}
    \includegraphics[width=\columnwidth]{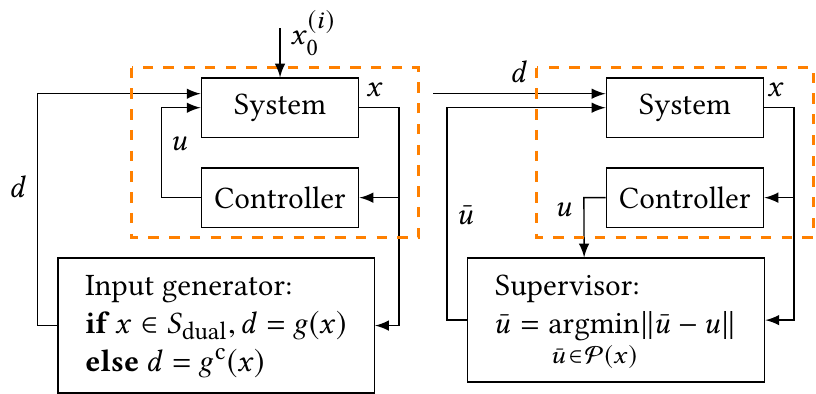}
  \end{center}
  \caption{In the evaluation phase, a (known) system model is controlled by a (black-box) controller. We discuss two settings, i.e., falsification (left) and supervision (right), for analyzing and enforcing safety of the closed-loop system respectively. In falsification, the outputs of the framework in Figure \ref{fig:workflow} are used to guide exploration of initial conditions ($x_0^{(i)}$) and disturbances ($d$) that lead to safety violations. As a by-product, a supervisor architecture that enforces invariance by rejecting potentially unsafe inputs ($u$) can be added around a controller that is found unsafe in the falsification step.}
  \label{fig:block-diagram}
\end{figure}

Our goal in this paper is to search for interesting corner cases for falsification of closed-loop control systems. By interesting corner case, we mean a pair of initial condition and external (disturbance) input signal that leads to a trajectory violating a given safety specification, together with a certificate that it is possible to satisfy the specification for this initial condition and external input; therefore violation is indeed avoidable. We summarize the proposed framework in Figures \ref{fig:workflow} and \ref{fig:block-diagram}  before detailing the different components.  

\subsection{Controlled Invariant Sets}

Invariance properties are the most basic safety properties where the goal is to avoid an unsafe set at all times, and has been widely studied in the literature \cite{blanchini1999survey}. The maximal (robust) controlled invariant set is the set of all states inside the safe set from which there exists a controller that can guarantee remaining safe for all future times (under all possible realizations of uncertainty and disturbances). 

Formally, we define a controlled invariant set for a continuous-time system using a tangent cone. 
Let $S$ be a set in $\mathbb{R}^n$;  
a vector $y\in \mathbb{R}^n$ is called a feasible direction of set $S$ at $x\in S$ if there exists $\varepsilon >0$ such that $x + \delta y \in S$ for all $\delta \leq \varepsilon$.
The tangent cone of a set $S$ at $x$ is then defined to be $T_S(x): = \text{closure}(\{y\mid y \text{ is feasible direction of }S \text{ at }x\})$.
Consider a dynamical system described by the following differential equation 
\begin{align}
\tfrac{d}{dt}x = f(x,u,d),
\label{eq:sysC}
\end{align}
where $x\in X$ is the state, $u\in U$ is the control, and $d\in D$ is the disturbance. Here, $X$, $U$, and $D$ represent the set of possible states, controls, and disturbances, respectively.
Set $S_{\rm inv}\subseteq X$ is called controlled invariant under the dynamics in Eq. \eqref{eq:sysC} if \cite{blanchini1999survey} 
\begin{align}
\forall x\in \partial S_{\rm inv}: \exists u \in U: \forall  d\in D: f(x,u,d)\in T_{S_{\rm inv}}(x),
\label{eq:CInvCond1}
\end{align}
where $\partial S_{\rm inv}$ represents the boundary of set $S_{\rm inv}$.
Set invariance can be defined similarly for discrete-time control systems of the form
\begin{align}
x(t+1) = F\big(x(t), u(t), d(t)\big).
\label{eq:sysD}
\end{align}
A set $S_{\rm inv}$ is controlled invariant under dynamics in Eq. \eqref{eq:sysD} if 
\begin{align}
\forall x\in S_{\rm inv}: \exists u \in U: \forall  d\in D: F(x,u,d)\in S_{\rm inv}.
\label{eq:CInvCond2}
\end{align}

For simple linear system dynamics subject to additive disturbance or polytopic uncertainty, it is possible to approximate the maximal invariant set to an arbitrary precision \cite{DeSantis:2004et,rungger2017computing}. In this paper, we used polytopic invariant sets, as is done in \cite{smith2016interdependence,nilsson2016correct,pn_thesis2017}. It is also possible to compute controlled invariant sets represented via barrier functions using sum-of-squares optimization \cite{prajna2004safety} or approximate them via abstraction-based techniques \cite{roy2011pessoa}. Invariant set computation can be seen as a safety game between the control input $u$ and the disturbance $d$, where the maximal controlled invariant set corresponds to the winning set (i.e., the set of all the initial states from which $u$ can enforce safety irrespective of the values of $d$) in the game for the control input.

\subsubsection{Supervision}
The controlled invariant set can be used to supervise a legacy controller to avoid violation of the safety constraint \cite{nilsson2016correct}, even when a controller for which a safety violation is found in the falsification step is used.
The idea is to provide a recursive guarantee on safety by enforcing the trajectory to stay within a controlled invariant set $S_{\rm inv}$ contained by the safe set. 
The supervisor is a set-valued map $\mathcal{P}$ that maps the current state $x_{\rm c}$ to a set of control inputs $\mathcal{P}(x_{\rm c})\subseteq  U$. 
Under any control $u\in \mathcal{P}(x_{\rm c})$, the next state stays within set $S_{\rm inv}$ under all disturbance.
The supervisor overrides the legacy controller in a minimally intrusive way. 
That is, the supervisor is active and provides a control input in set $\mathcal{P}(x_{\rm c})$, whenever the legacy controller gives a control input outside $\mathcal{P}(x_{\rm c})$ at state $x_{\rm c}$. As shown in Fig. \ref{fig:block-diagram}, when the legacy controller's input $u$ is in $\mathcal{P}(x_{\rm c})$, we have the supervisor output $\bar{u} = u$.

To be specific, the set $\mathcal{P}(x_{\rm c})$ can be constructed in the following way.
Let $x(t+1) = F\big(x(t),u(t),d(t)\big)$ be the discrete-time dynamics. We define the set 
\begin{align}
\mathcal{P} := \{(x,u)\mid F(x,u,d)\in S_{\rm inv}, \ \forall d\in D\}.
\end{align}
Given the current state $x_{\rm c}$, set $\mathcal{P}(x_{\rm c})$ 
is obtained by fixing the $x$ component of the points in $\mathcal{P}$ to be $x_{\rm c}$, i.e., $\mathcal{P}(x_{\rm c}) := \{(x,u)\in \mathcal{P}\mid x = x_{\rm c}\}$. 
In particular, under the assumption that $S_{\rm inv}$ is a polyhedron (or a union of polyhedra, resp.), $F$ is linear in $x$, $u$, $d$, and $D$ is a polyhedron, then $\mathcal{P}$ can also be represented as a  polyhedron (or a union of polyhedra, resp.).

\subsection{Sampling of the Boundary}\label{sec:sampl}

\begin{figure}[t] 
	\centering 
	\includegraphics[width=0.85\linewidth]{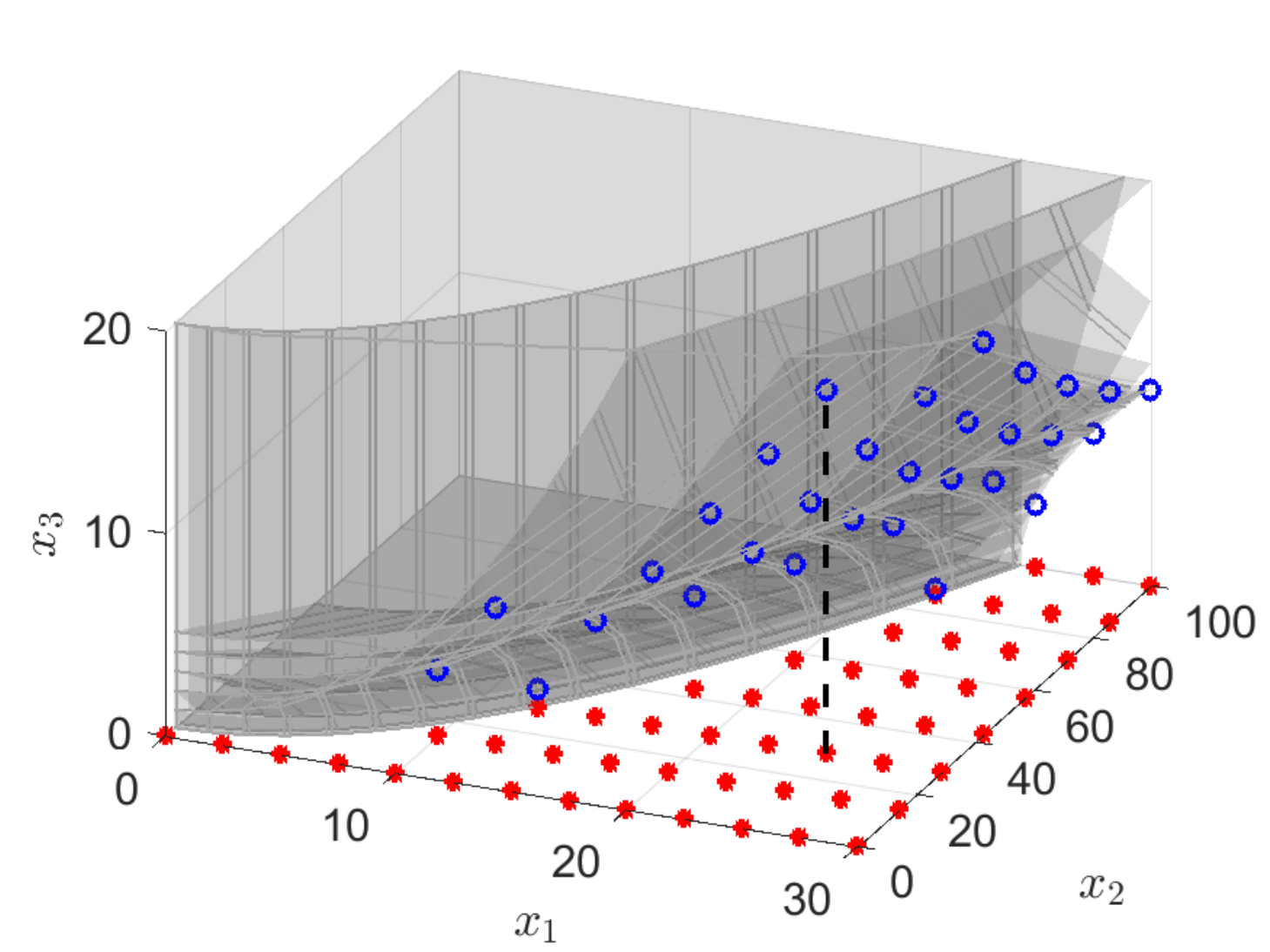}
	\caption{Sampling the boundary of a union of polyhedra.}\label{fig:sampleB} 
\end{figure}

We sample the boundary of the controlled invariant set to obtain potentially interesting initial conditions. As mentioned before, the focus in this paper is on controlled invariant sets that can  be represented as a finite union of polyhedra.
Figure \ref{fig:sampleB} illustrates the boundary sampling scheme of a polyhedron-union set.
The gray shaded area is the union of the polyhedra. 
We assume the union set is contained within a hyper-rectangular domain, and sample along the first $n-1$ dimensions of the domain. 
These samples corresponds to the red dots in the figure.
Then the red dots are projected onto the boundary of the invariant set, which are marked by the blue circles in the figure.
In particular, this projection can be done by the following procedure.
\begin{enumerate}
\item[1)] We first project each red dot $y$ onto the boundary of each polyhedron $P = \{x\in\mathbb{R}^n\mid Ax\leq b\}$ in the collection. To be specific, we fix the first $n-1$ coordinates of points $x$ inside polyhedron $P$ to be the same as the red dot $y$.
This gives 1 dimensional polyhedron
\begin{align}
P_y := \{x\in\mathbb{R}^n\mid Ax\leq b, \ x_i = y_i, i = 1\dots,n-1\}.
\end{align}
We then compute the vertex representation of the set $P_y$ using MPT3 \cite{herceg2013multi}. 
\item[2)] The vertices of $P_y$ are admitted if they are not in the interior of other polyhedra.
\end{enumerate}
\begin{remark}
The proposed sampling scheme can be extended to the case where the controlled invariant set is represented as the union of convex sets in form of $C = \{x\in \mathbb{R}^n \mid f_j(x)\leq 0, j = 1,\dots,m\}$. 
Similar to step 1) in the above procedure, set $C_y$ is created as
\begin{align}
C_y := \{x\in\mathbb{R}^n\mid & f_j(x)\leq 0, \  j = 1,\dots, m\nonumber \\
& x_i = y_i, \ i = 1\dots,n-1\}.
\end{align}
Set $C_y$ is a 1-D interval whose bounds can be computed by solving 1-D convex optimization problems $\min \{x_n\mid x\in C_y\}$ and $\min \{-x_n\mid x\in C_y\}$. Also note that a union of convex sets is not necessarily convex and may contain holes. Our proposed approach is able to sample the boundary of the holes as well.
\end{remark}

For other types of sets, there is a brief survey in \cite{ghosh2013nearly} on the existing approaches to sample the surface of nonconvex polyhedra.
Other methods for generating (asymptotically) uniform samples on a polytope's boundary include the shake-and-bake method \cite{smith1984pointsuniform,boender1991shakeandbake}, and sweep plane method \cite{Leydold98asweep-plane}, and these can be used as alternatives to the approach described above.

\subsection{Computing the Falsifying Inputs}\label{sec:inverse_game}

\subsubsection{Dual Game}
A falsifying scenario consists of two parts: an initial condition and a disturbance input profile.  In this part, we show how to compute a falsifying input profile, through solving the so called dual game, given that the initial condition is outside the maximal invariant set. 
Theoretically, if the initial state is already outside the maximal invariant set, there exists a disturbance input profile that steers the trajectory outside the safe set. However, if the disturbance profile is not selected carefully, it does not necessarily lead to falsification.
\begin{figure}[t] 
	\centering 
	\includegraphics[width=0.6\linewidth]{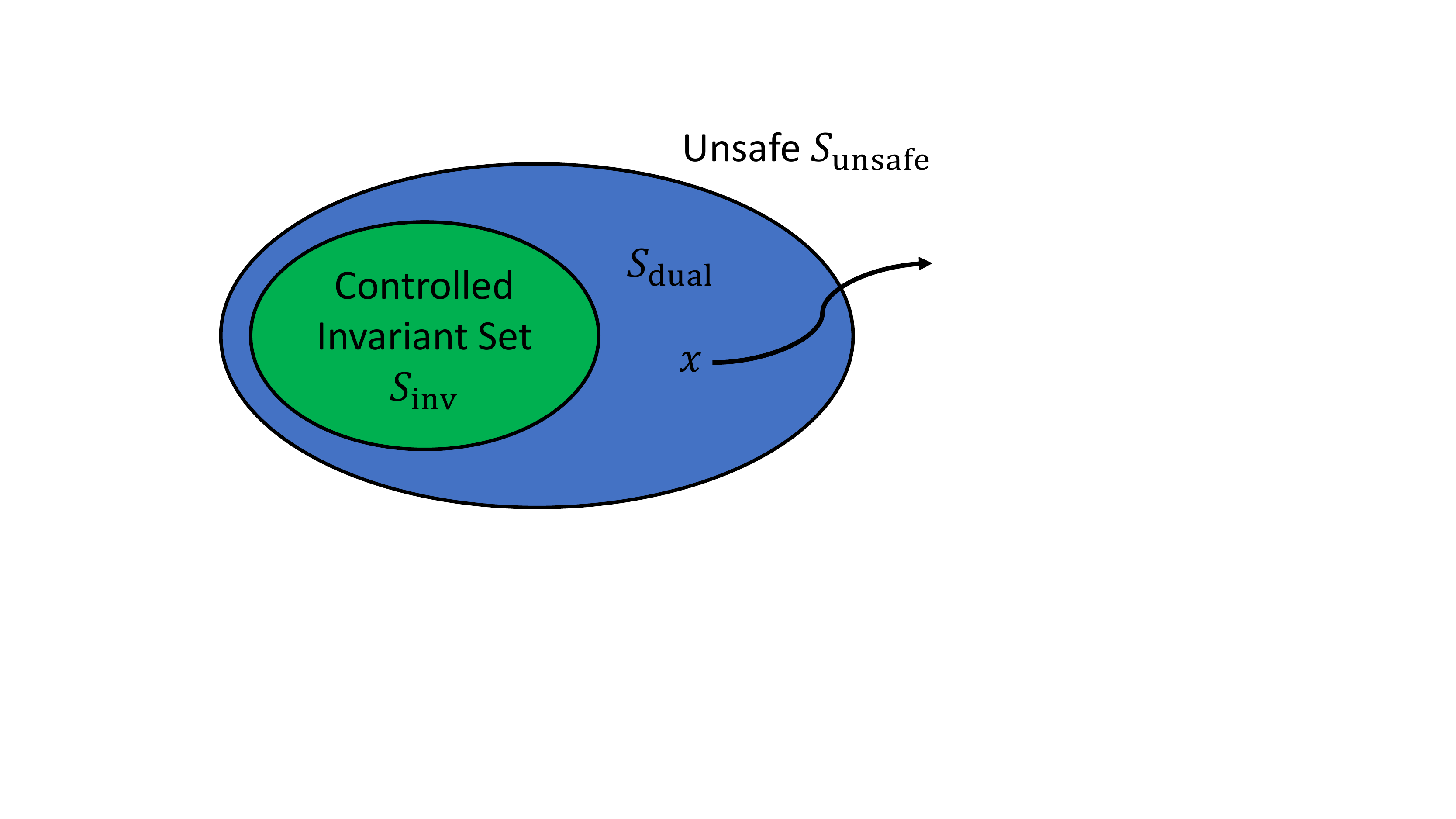}
	\caption{Illustration: objective of the dual game}\label{fig:dual_game} 
\end{figure}

We first define some terminology. Let the system dynamics be given by Eq. \eqref{eq:sysC}, and let $S_{\textrm{safe}}$ be the safe set we want to stay inside for all time. The invariance game aims at finding the largest controlled invariant set $S_{\textrm{inv}}\subseteq S_{\textrm{safe}}$. Figure \ref{fig:dual_game} shows the objective of its dual game: we want to find set $S_{\textrm{dual}}$,  and a dual strategy $g: S_{\textrm{dual}}\rightarrow D$, under which the states starting from $S_{\textrm{dual}}$ is steered into unsafe set $S_{\textrm{unsafe}}: = (S_{\textrm{safe}})^C$ in finite time, as long as $u\in U$.

We solve the dual game by computing the backwards reachable set of unsafe set $S_{\textrm{unsafe}}$. 
For linear discrete-time dynamics, assuming that unsafe set $S_{\textrm{unsafe}}$ is a polytope, the backwards reachable set can be computed as a collection of polytopes using the same approach in \cite{nilsson2016correct}. 
The only difference is that we are now ``controlling'' the disturbance $d$ and trying to be robust to the real control action $u\in U$.
To be specific, let the dynamics be 
\begin{align}\label{eq:dyn_inputgen}
x(t+1) = Ax(t) + Bu(t) + Ed(t) + K
\end{align}
where $x\in X$, $u\in U$, $d\in D$ are polytopes. 
We first compute a sequence of polytopes, starting with $P_0 = S_{\rm unsafe}$, as follows: 
\begin{align}
P_{i+1} = & \{(x,d)\in X\times D \mid \forall u\in U: \nonumber \\
 & Ax + Bu + Ed + K \in P_i \} \end{align}
We then project each polytope $P_i$ onto $X$ space to obtain $\{\overline{P}_i\}$, and the winning set of the dual game is given by $\bigcup_i \overline{P}_{i}$.
To determine the dual game strategy $g$ at the current state $x$, we locate $x$ in one of the projected polytopes $\overline{P}_i$, and the dual strategy can be generated by picking $g(x)$ such that $(x,g(x))\in P_i$. 

When $S_{\rm unsafe}$ is nonconvex but can be expressed as a union of polytopes, we compute the backwards reachable set for each polytope and take the union of the obtained backwards reachable sets. This gives a conservative, yet sound, winning set for the dual game. Note that, when the invariant set or the winning set for the dual game is computed via such a conservative approach, there will be a gap between $S_{\textrm{inv}}$ and $S_{\textrm{dual}}$ in Fig.~\ref{fig:dual_game}, corresponding to a set of initial conditions for which concluding whether they are ``interesting" or not is not possible with the computed sets.

\subsubsection{Ellipsoid Method}
Note that the dual strategy $g$ is defined on $S_{\textrm{dual}}$ and is not applicable everywhere on $S_{\textrm{safe}}$. Thus we need a \emph{complementary strategy} $g^c$ to generate falsifying inputs for states $x \in S_{\textrm{safe}}\setminus S_{\textrm{dual}}$. Next, we propose some heuristics for computing a complementary strategy. Assume that the safe set is given as a union of polyhedra $S_{\textrm{safe}} = \cup_i S_{\textrm{safe},i}$ and $C_{\textrm{safe}}$ denotes the convex-hull of $S_{\textrm{safe}}$. It is shown in \cite{john2014extremum} that, for any compact set $C$, there exists a unique minimum volume ellipsoid (called \emph{LJ-ellipsoid} of $C$) covering it.
Denote the LJ-ellipsoid of $C_\textrm{{safe}}$ with $E_{\textrm{safe}}$, which is defined as 
\begin{equation}
E_\textrm{{safe}} = \{ x\in X \mid [x^\top, 1] Q [x^\top, 1]^\top \leq 1, Q>0\},
\end{equation}
where the positive definite matrix $Q$ parametrizing the ellipsoid can be computed using \cite{rimon1992efficient}.
Define the \emph{level} of $x \in  X$ as 
\begin{equation}
l(E_\textrm{{safe}},x) = [x^\top, 1] Q [x^\top, 1]^\top.
\end{equation}
It is reasonable to assume that points lying on the higher levels are closer to the unsafe set; hence, driving the system to higher levels would force it either to the unsafe set $S_{\textrm{unsafe}}$ or to the winning set of the dual game  $S_{\textrm{dual}}$. With this intuition, complementary strategy $g^c:S_{\textrm{safe}}\setminus S_{\textrm{dual}} \to D$ is defined such that it steers the system to the highest possible level set at each step:
\begin{equation}\label{eq:comp_strategy}
g^c(x) \doteq \argmax_d \{l(E_\textrm{{safe}},x')\;|\; x' = Ax + Bu + Ed + K\},
\end{equation}
where we assume that control input $u$ is known.\footnote{When the invariant set is unbounded, LJ-ellipsoid does not exist. In this case $g^c$ can be computed by computing inputs that steer the state closer to $S_{\textrm{dual}}$ by directly minimizing the distance to $S_{\textrm{dual}}$ though the corresponding optimization problem can be more complex. Alternatively, if the rays corresponding to unbounded directions are known, an ellipsoid that is significantly elongated along those directions can be chosen by bounding those rays at a large enough level.}

Falsifying inputs are computed using $g$ if the current state $x\in S_\textrm{{dual}}$, and $g^c$ is used otherwise (see Fig.~\ref{fig:block-diagram}). Additionally, we develop some simple input-generation heuristics tailored for the ACC and LK functions of autonomous driving. These tailored heuristics will be presented on the fly in Section \ref{sec:CaseStudy}.

\section{System model and specifications}

For the case studies we consider two autonomous driving subsystems: adaptive cruise control and lane keeping. An adaptive cruise controller (ACC) controls the speed of the vehicle to follow a desired speed if there is no car in front, and to follow the lead vehicle within some safe following distance (headway) if there is a relatively slower lead vehicle in front. A lane keeping (LK) controller controls the steering of the vehicle to avoid lane departures. Therefore, adaptive cruise control controls the longitudinal dynamics and lane keeping control deals with the lateral dynamics. In the rest of this section, we provide dynamical models used in our examples and formalize safety specifications for both systems.

\subsection{Longitudinal and Lateral Dynamics}
We use the following model from \cite{nilsson2016correct} to 
describe the longitudinal dynamics of the vehicle:
\beq\label{eq:acc}
\frac{\mathrm{d}}{\mathrm{d}t} \begin{bmatrix} \vF \\ \headway \\ \vL \end{bmatrix} = \begin{bmatrix} 
\tfrac{1}{m}( \Fw - f_0 - f_1 \vF- f_2 \vF^2)\\
\vL - \vF\\ 
 \aL
\end{bmatrix}.
\eeq
The system states consist of the following car velocity $\vF$, lead car velocity $\vL$, and the headway $\headway$ (i.e., the relative distance between the lead and following car).
Control input $\Fw$ represents the net force acting on the mass of the following car.
The lead car acceleration $\aL$ can be viewed as a disturbance to the system.
Finally, constants $m$, $f_0$, $f_1$ and $f_2$ are parameters of the model. 
The values of these parameters and the bounds of the variables can be found in Table \ref{tab:params_acc}. In particular, the domain the dynamics are defined on is $X_{ACC} := [\vFmin,\vFmax]\times[\hmin,\infty)\times [\vLmin,\vLmax]$.
{\small
\begin{table}[h] 
	\centering 
	\caption{Parameter values for the ACC model}  \label{tab:params_acc}
	\begin{tabular}{lll}
		\textbf{Param.} & \textbf{Description} & \textbf{Value} \\ \hline
		$m$ & car+cargo mass  & 1462 ($kg$) \\
        $f_0$  & friction/drag term &  51 ($N$) \\
        $f_1$  & friction/drag term &  1.2567 ($Ns/m$) \\
	    $f_2$ & friction/drag term &  0.4342 ($Ns^2/m^2$) \\
\hline
        $\vFmin,\vLmin$  & minimal car velocity  & $0$ ($m/s$) \\ 
        $\vFmax,\vLmax$  & maximal car velocity  & 25 ($m/s$) \\ 
        $\vFdes$ & desired car velocity   & 20 ($m/s$) \\ 
        $\Fwmin$ & minimal force, comfort  &  $-4305.9$ ($N$) \\ 
        $\Fwmax$ & maximal force, comfort   &  2870.6 ($N$) \\ 
        $\Fwminphys$ & minimal force, physical  &  $-11482.5$ ($N$) \\ 
        $\Fwminphys$ & minimal force, physical  &  $7176.6$ ($N$) \\ 
        $\aLmin$ & minimal acceleration  & $-0.97$ ($m/s^2$) \\ 
        $\aLmax$ & maximal acceleration   & 0.65 ($m/s^2$) \\ 
        $\wmin$ & minimal time headway & 1.7 ($s$)\\
	    $\hmin$ & minimal headway  & 4 ($m$) \\
		 \hline
	\end{tabular}
\end{table} }

The lateral dynamics are described by 
\begin{equation}\label{eq:lk}
\begin{array}{l}
\frac{\mathrm{d}}{\mathrm{d}t} \underbrace{\begin{bmatrix} \ldev \\ \vLateral \\ \adev \\ \yaw \end{bmatrix}}_{\xLK}
=  \underbrace{\begin{bmatrix} 0 & 1 & \vN & 0 \\
0 & -\frac{ C_{\alpha f} + C_{\alpha \yaw} }{ m\vN } & 0 & \frac{b C_{\alpha \yaw} - a C_{\alpha f} }{ m\vN } - \vN \\
0 & 0 & 0 & 1  \\
0 & \frac{b C_{\alpha \yaw} - a C_{\alpha f} }{ I_{z} \vN } & 0 & -\frac{a^2 C_{\alpha f} + b^2 C_{\alpha r} }{ I_{z} \vN }
\end{bmatrix}}_{A_{LK}}
\begin{bmatrix} \ldev \\ \vLateral \\ \adev \\ \yaw \end{bmatrix}\\
\quad\quad\quad\quad\quad + 
\underbrace{\begin{bmatrix} 0 \\ \frac{C_{\alpha f} }{m} \\ 0 \\ a \frac{C_{\alpha f} }{I_{z}} \end{bmatrix}}_{B_{LK}} \delta_f
+  \begin{bmatrix} 0 \\ 0 \\ -1 \\ 0 \end{bmatrix} r_{d},
\end{array}
\end{equation}
where the states are: lateral deviation from the center of the lane ($\ldev$), the lateral velocity ($\vLateral$), the yaw-angle deviation in road-fixed coordinates ($\adev$), and the yaw rate ($\yaw$), respectively. The input $\delta_f$ is the steering angle of the front wheels, which is limited to lie within $\theta_s^{\rm min}$ and $\theta_s^{\rm max}$; and $r_d$ is the desired yaw rate, which we interpret as a time-varying external disturbance and computed from road
curvature by $r_d = v/R_0$ where $R_0$ is the (signed) radius of
the road curvature and $v$ is the vehicle's longitudinal velocity.
Other parameters include $m$, the total mass of the vehicle, and $a$, $b$, $C_{\alpha f}$, and $C_{\alpha r}$, which are vehicle geometry and tire parameters. All values can be found in Table \ref{tab:params_lk}. Accordingly, the domain the dynamics are defined on is $X_{LK} := [-y^{\rm max},y^{\rm max}]\times[-\nu^{\rm max},\nu^{\rm max}]\times [-\Delta\Psi^{\rm max},\Delta\Psi^{\rm max}]\times[-r^{\rm max},r^{\rm max}]$.

{\small
\begin{table}[h] 
	\centering 
	\caption{Parameter values for the LK model}  \label{tab:params_lk}
	\begin{tabular}{lll}
		\textbf{Param.} & \textbf{Description} & \textbf{Value} \\ \hline
        $\vN$  & nominal velocity  & 20 ($m/s$) \\ 
        $m$  & car+cargo mass  & 1462 ($kg$) \\ 
        $I_z$ & car moment of inertia   & 2500 ($kg m^2$) \\ 
        $a$ & vehicle geometry parameter  & 1.08 ($m$) \\ 
        $b$ & vehicle geometry parameter   & 1.62 ($m$) \\ 
        $C_{\alpha f}$ & tire parameter  & 85400 ($N/rad$) \\ 
        $C_{\alpha r}$ & tire parameter   & 90000 ($N/rad$) \\ \hline
        $y^{\rm max}$ & maximum lateral deviation & 0.9 (m)\\
        $\nu^{\rm max}$ & maximum lateral velocity    &  1 (m/s) \\ 
        $\Delta\Psi^{\rm max}$ & maximum yaw-angle deviation    &  0.15 (rad) \\
        $r^{\rm max}$ & maximum yaw rate   &  0.27 (rad/s) \\ 
        $\theta_s^{\rm min}$ & minimum steering angle & $-0.26$ (rad)\\
        $\theta_s^{\rm max}$ & maximum steering angle& 0.26 (rad)\\
		 \hline
	\end{tabular}
\end{table} }

\subsection{Formal Specifications for ACC and LK}

For ACC, we focus on the safety aspect of requirement in this work. The (safety part of) ISO Standard requirements for Adaptive Cruise Control Systems \cite{isoACCstandard} state:
\begin{enumerate}
\item the control input should stay within specified bounds all the times.
\item 
whenever the lead car is close in the sense that the headway $\headway < \vFdes\wdes$, the time headway $\omega$ needs to satisfy $\omega \geq \wmin$ at all times.
\end{enumerate}

We extract the safety part of the above ISO requirement and express it formally in logic. 
Define sets
\begin{align}
\begin{array}{l}
    M := \{ (\vF, \headway, \vL) \mid \vFdes > \headway/\wdes \}, \\
    S := \{ (\vF, \headway, \vL) \mid \vF \leq \headway/\wmin, \headway \geq  \hmin \}, \\
    S_U := \{ \Fw \mid \Fwmin \leq \Fw \leq \Fwmax \}.
\end{array}
\end{align}
Set $M$ is the set of states where the lead car is close, set $S$ is the safe set of states, and set $S_U$ contains the allowable control inputs.  Adding the speed limits encoded by the domain $X_{ACC}$, the overall specification can be expressed as
\begin{align}\label{spec:ACC_original}
\nonumber \left(\forall t: \Fw(t) \in S_U\right) \wedge \Big(\forall t: \big((\vF(t), \headway(t), \vL(t)) \in M\big)\rightarrow \\
 \big((\vF(t), \headway(t), \vL(t)) \in S\cap X_{ACC}\big)\Big).
\end{align}
To check safety in the presence of a close enough lead car, we assume the states are in $M$ and consider the following safety specification, denoted by $\varphi_{ACC}$:
\begin{align}\label{spec:ACC}
\left(\forall t: \Fw(t) \in S_U\right) \wedge \Big(\forall t: \big((\vF(t), \headway(t), \vL(t)) \in S\cap X_{ACC}\big)\Big).
\end{align}
In the later falsification experiments, we will consider violations of different aspects of the specification $\varphi_{ACC}$, that is, 
\begin{align}
\varphi_{ACC}^1&:=  \forall t: \vF(t) \leq  \headway/\wmin, \nonumber\\
\varphi_{ACC}^2&:=  \forall t:  \headway(t)\geq \hmin, \label{eq:ACC_types}\\ 
\varphi_{ACC}^3&:=  \forall t: h(t) \geq 0. \nonumber
\end{align}
These three safety specifications correspond to small time headway, small distance headway, and crash, respectively. 
Note that  specification $\varphi_{ACC}^2$ implies $\varphi_{ACC}^3$ as $\hmin > 0$. Here we distinguish specification $\varphi_{ACC}^3$ from $\varphi_{ACC}^2$ because violating $\varphi_{ACC}^3$ is considered to be more severe.

For LK, as mandated by the width of roads in the United States (approx. $3.8$m) and typical car widths (approx. $2$m), the specification states that the car must stay within $y^{\rm max}$ meters of the center of the lane, i.e. $|y(t)| \le y^{\rm max}$. We also require the other states to remain in the domain $X_{LK}$ as larger values of these states are either physically less meaningful (e.g., can correspond to the vehicle navigating in the reverse direction) or violate passenger comfort requirements. Moreover, the lateral dynamics model we use is valid for relatively smaller ranges of yaw rate, yaw angle and lateral velocity. With these requirements, the overall specification for LK, denoted by $\varphi_{LK}$, is formally stated as: 
\begin{align}
\forall t: &  \big( y(t), \nu(t),\Delta\Psi(t),r(t) \big)\in X_{LK}. \label{eq:LK_type2}
\end{align}
Note that state $y$ being out of bound should be considered to be a significant safety violation, while the other three states in $x_{LK}$ being out of bounds leads to a less comfortable ride.  
Therefore, we will independently count the violations of the specification below in the falsification experiments:
\begin{equation}\label{spec:LK}
\varphi_{LK}^1:= \forall t: |y(t)| \le y^{\rm max}.
\end{equation}

\section{Case studies}\label{sec:CaseStudy}

In this section, we evaluate the proposed approach on case studies with different controllers for adaptive cruise control and lane keeping. 

In what follows, both the wheel force $F_w$ and the steering angle $\delta_f$ are bounded quantities. Thus, for controllers that cannot handle such input constraints, we use a saturation function before feeding their output to the system. The saturation function $sat$ is defined as follows:
\begin{align}
sat_{\underline{x}}^{\overline{x}}(x) = 
\begin{cases}
\underline{x} & \text{ if } x \leq \underline{x} \\
x & \text{ if } \underline{x} < x < \overline{x}\\
\overline{x} & \text{ if } x\geq \overline{x} \\
\end{cases}.
\end{align} 

In addition to sampling at the boundary of the invariant set, we also sample in the interior of the invariant set
to generate less ``trickier" initial conditions. These interior points are  obtained by shifting (for ACC) or scaling (for LK) the boundary samples.

\subsection{Adaptive Cruise Control Results}
We computed a controlled invariant set $S_{ACC}$ for the longitudinal dynamics in \eqref{eq:acc}, and sampled the boundary of this set with the proposed approach to find falsifying initial conditions. 
The disturbance profile is computed by (i) solving the dual game, (ii) a simple heuristic that corresponds to the lead car doing a maximum braking, or (iii) the lead car trying  to achieve $v^{\rm des}$. 
We explored the following three classes of controllers for meeting the ACC requirement. 
\begin{enumerate}
\item For the first controller, we performed feedback linearization followed by pole placement with a hybrid proportional (P) controller, defined as: 
\begin{align}
u = f_0  + f_2\vF^2 - k_P\big(v - \min(\vFdes, \headway/\wdes)\big),
\end{align}
where $k_P$ is the proportional gain, the $\min$ part takes care of the two different ACC modes and $F_w =  sat_{\Fwmin}^{\Fwmax}(u)$ given the input saturations.  

\item We also consider hybrid proportional-integral (PI) controllers with the following dynamics:
\begin{align}
u(t) = & f_0  + f_2\vF^2 - k_P\big(v(t) - \min(\vFdes, \headway(t)/\wdes)\big) \nonumber \\
& - k_I e(t),\\
e(t) = & \sum_{\tau=0}^t \big(v(\tau) - \min(\vFdes, \headway(\tau)/\wdes)\big), 
\end{align}
where $e(t)$ is the error state and $k_I$ is the integral coefficient. 
Similarly, the control input $u$ needs to be saturated to obtain practical $\Fw$.  
\item We also designed an MPC controller with 
a linearized discrete-time model with a sampling period of $0.1$s using the following formulation:
\begin{align}
\begin{array}{rl}
\min & \sum_{t = 0}^T\Vert\vF(t) - \min(\vFdes,\headway(t)/\wdes)\Vert\\
\text{s.t.}& \text{Linearized, time-discretized dynamics of ACC,}\\
& \Fwmin \leq \Fw(t) \leq \Fwmax, \ \ t = 0,\dots,T-1,\\
& \vLmin\leq \vL(t)\leq \vLmax, \ \ t = 0,\dots,T,\\
& \vFmin\leq \vF(t)\leq \vFmax, \ \ t = 0,\dots,T,\\
& 0 \leq \headway(t), \ \ t = 0,\dots,T.\\
& \vF(0) = v_0, \headway(t) = h_0, \vL(t) = v_{L,0} 
\end{array}
\end{align}
where $v_0$, $h_0$, $v_{L,0}$ are the initial conditions and $T$ is the length of the prediction horizon. 
Since the objective contains term $\min(\vFdes,\headway(t)/\wdes)$, the MPC is hybrid in its nature. 
To simplify the computation load, we replace the target velocity throughout the predicting horizon by $\min(\vFdes,\headway(0)/\wdes)$, so that the MPC problem can be solved by a QP solver. 
\end{enumerate}
Tables \ref{tab:ACC_PPiMpc_noinput}-\ref{tab:ACC_PPiMpc_accelerate} summarize the falsification rates for the samples both from the interior and the boundary of the controlled invariant set $S_{ACC}$, with disturbance $\aL$ profile generated by multiple methods.
It should be noted that the same controlled invariant set $S_{ACC}$ is used to generate initial states to investigate the three different aspects of the safety specification in Eq. \eqref{eq:ACC_types}. This is because set $S_{ACC}$ is synthesized against the overall safety specification in Eq. \eqref{spec:ACC}. 
Overall, the MPC controller seems better than the naively designed P controller in terms of safety.

Another key observation  is that the falsification rates of the test cases with interior initial conditions can be higher or lower than those on the boundary of set $S_{ACC}$, depending on how the disturbance (i.e., $a_L$) profile is generated. 
In particular, the test cases with interior initial conditions have higher falsification rate than the boundary cases in Table \ref{tab:ACC_PPiMpc_noinput}-\ref{tab:ACC_PPiMpc_maxbrake}, and usually have lower falsification rate in Table \ref{tab:ACC_PPiMpc_accelerate}.
The key difference between Table \ref{tab:ACC_PPiMpc_noinput}-\ref{tab:ACC_PPiMpc_maxbrake} and Table \ref{tab:ACC_PPiMpc_accelerate} is that the leading car is usually decelerating (or maintaining speed constant) in the test cases from Table \ref{tab:ACC_PPiMpc_noinput}-\ref{tab:ACC_PPiMpc_maxbrake}, while it is accelerating under the test cases from Table \ref{tab:ACC_PPiMpc_accelerate}. In what follows we briefly discuss how this difference affects the falsification rates by the interior initial conditions and by the boundary ones.

When the lead car is decelerating, the dynamics tend to have a ``steady state'' outside the controlled invariant set $S_{ACC}$. This is true because the lead car's deceleration shortens the headway $h$ and hence  pushes the state towards the boundary of $S_{ACC}$. In this case, the falsifications are due to the long term behavior of the dynamics as a trajectory may eventually leave $S_{ACC}$ Since such undesired behaviors occur in a longer term, starting from the interior of set $S_{ACC}$ may not prevent ultimate falsification. 

Moreover, the trajectories initiating from the interior tend to move to the ``trickier" parts of the boundary, where safe actions are limited, which increases the falsification rate. We next explain why this is so. 
First note that a point in the interior of $S_{ACC}$ usually has larger relative headway $\headway$ and larger lead car velocity $\vF$. 
Consequently, the target velocity defined by $\min(\vFdes, \headway/\wdes)$ has a higher chance to be equal to $\vFdes$ when starting from the interior.
The controller hence accelerates to achieve $\vFdes$ and maintains the velocity there. 
Now since the lead car velocity $\vL$ is low (due to deceleration or small initial value), such acceleration will eventually lead to small headway $\headway$, which will change the target velocity from $\vFdes$ to $\headway/\wdes$. At that moment, however, the following car velocity may be already relatively high. This hence leads to a harder scenario and increases the chance of falsification, which explains the result in Table \ref{tab:ACC_PPiMpc_noinput}-\ref{tab:ACC_PPiMpc_maxbrake}.

On the contrary, when lead car's steady state speed is $\vFdes$, i.e. it is mostly accelerating, the dynamics tend to have a ``steady state'' inside the set $S_{ACC}$. This is true because the lead car's acceleration enlarges headway $h$ and pushes the state towards inside of $S_{ACC}$. 
In this case, the falsifications are mainly due to the transient state of the dynamics because the state will eventually converge to that steady state inside $S_{ACC}$.
By our conjecture, the initial conditions on the boundary of $S_{ACC}$ have higher chances for capturing the falsifications due to transient state. This explains why the falsification rate in Table \ref{tab:ACC_PPiMpc_accelerate} agrees with our conjecture.

To summarize, initial conditions on the boundary help identify safety violations in the transient behavior, whereas input generation techniques tend to capture safety violations due to persistent disturbances (i.e., ``steady state").

\subsection{Lane Keeping Results}

Let us denote the state vector $[\ldev, \vLateral, \adev, \yaw]^\top$ in Eq. \eqref{eq:lk} as $\xLK$. We computed a controlled invariant set $S_{LK}$ for the lane keeping model in \eqref{eq:lk}, sampled the boundary of this set with the proposed approach to find falsifying initial conditions, and use various input generation methods to generate road profiles. We explored three classes of controllers for meeting the lane keeping requirement:
\begin{enumerate}
\item A proportional (P) state feedback controller, defined as 
	\begin{equation}u =     K_P^\top \xLK.\end{equation}
    Several P controllers are designed by placing the poles in different locations. The control input $\delta_f$ is obtained by saturating $u$ to account for the practical limit of the actuator, i.e.,  $\delta_f = sat_{\theta_s^{\rm min}}^{ \theta_s^{\rm max}}(u)$. \item A proportional-integral (PI) controller, defined as 
\begin{equation} 
u = K_I^\top \widetilde{x}_{LK}
\end{equation}
where $\widetilde{x}_{LK}$ expands $\xLK$ to include an error state $e$: 
	\begin{equation}\frac{d}{dt}{\begin{bmatrix}
         \xLK \\
         e
        \end{bmatrix}} = \begin{bmatrix}
         A_{LK} & \mathbf{0}_{4\times1} \\
         [1\;0\;0\;0] & 0 \\
        \end{bmatrix} \begin{bmatrix}
         \xLK \\
         e
        \end{bmatrix} + \begin{bmatrix}
        B_{LK} \\
         0
        \end{bmatrix} u. \end{equation}
Several PI controllers are designed by choosing different pole locations. Similarly, the control input $\delta_f$ is obtained by saturating $u$ accordingly. \item An MPC controller with the following formulation:
\begin{align}
\begin{array}{rl}
\min & \sum_{t = 0}^T \xLK(t)^\top Q\xLK(t) + u^2(t)\\
\text{s.t.} & \text{Time-discretized dynamics of LK},\\
& \theta_s^{\rm min} \le u(t) \le \theta_s^{\rm max}, \ t = 0,\dots, T-1\\
& \xLK(0) = x_0.
\end{array}
\end{align}
Since the input saturation is accounted for by the constraints in the MPC formulation, control input $\delta_f = u$. 
where $x_0$ is the initial condition, and $Q = \diag([1 \ 0 \ 0 \ 0])$. 
\end{enumerate}
Table \ref{tab:LK_PPiMpc_dual_h} summarizes the falsification rates for the above three controllers. The initial conditions are generated by sampling the interior and the boundary of set $S_{LK}$, and the disturbance profiles are generated using ellipsoid method plus dual game and using a heuristic  described as follows: 
\begin{align}r_d = 
\begin{cases}
r_d^{\rm min} & \text{ if } y(t+\tau) \geq t(t) \\
r_d^{\rm max} & \text{ if } y(t+\tau) < t(t)\\
\end{cases},
\label{eq:LK_h}
\end{align}
where $\tau$ is the sampling time of the discrete-time system.

Overall, our MPC design seems safer than the PI design, which is safer than the P controller. Note that none of these designs are tuned properly, the goal is just to demonstrate how controlled invariant sets can be used to evaluate different designs.

\begin{figure*}[t]
	\centering 
	\includegraphics[width=0.9\linewidth]{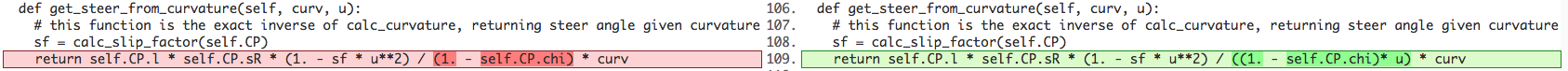}
	\caption{The output of a diff utility showing the modification in \comma{}. Left: \commastar{}, right: \comma{}.}\label{fig:bug_fix} 
\end{figure*} 

{\small
\begin{table}[!thpb]
\caption{Controllers used in our tests}
\label{tab:contrller}
\begin{center}
\begin{tabular}{l|c|c|c|}
\cline{2-4}
                                                    & \textbf{\begin{tabular}[c]{@{}c@{}}Controller\\ (parameters)\end{tabular}}                                 & \textbf{Notation} & \textbf{Parameter} \\ \hline\hline
\multicolumn{1}{|l|}{\multirow{9}{*}{\textbf{ACC}}} & \multirow{3}{*}{{\begin{tabular}[c]{@{}c@{}}Proportional\\ controller\\ (P gain)\end{tabular}}}                     & ${\rm P_{ACC}\ \#1}$         & $k_P = 600$                   \\ \cline{3-4} 
\multicolumn{1}{|l|}{}                              &                                                                                                                & ${\rm P_{ACC}\ \#2}$         &     $k_P = 1800$                \\ \cline{3-4} 
\multicolumn{1}{|l|}{}                              &                                                                                                                & ${\rm P_{ACC}\ \#3}$        &      $k_P = 4000$               \\ \cline{2-4} 
\multicolumn{1}{|l|}{}                              & \multirow{3}{*}{{\begin{tabular}[c]{@{}c@{}}PI\\ controller\\ (P/I gains)\end{tabular}}} & ${\rm PI_{ACC}\ \#1}$          &   $k_P = 600$,\;  $k_I$ = 200                 \\ \cline{3-4} 
\multicolumn{1}{|l|}{}                              &                                                                                                                & ${\rm PI_{ACC}\ \#2}$          &  $k_P=1800,\; k_I = 400  $                \\ \cline{3-4} 
\multicolumn{1}{|l|}{}                              &                                                                                                                & ${\rm PI_{ACC}\ \#3}$        &   $k_P = 4000,\;  k_I =2000$                \\ \cline{2-4} 
\multicolumn{1}{|l|}{}                              & \multirow{3}{*}{{\begin{tabular}[c]{@{}c@{}}MPC\\ (horizon)\end{tabular}}}                              & ${\rm MPC_{ACC}\ \#1}$          &2                    \\ \cline{3-4} 
\multicolumn{1}{|l|}{}                              &                                                                                                                & ${\rm MPC_{ACC}\ \#2}$         &   8                 \\ \cline{3-4} 
\multicolumn{1}{|l|}{}                              &                                                                                                                & ${\rm MPC_{ACC}\ \#3}$          &   20                 \\ \hline\hline
\multicolumn{1}{|l|}{\multirow{9}{*}{\textbf{LK}}} & \multirow{3}{*}{{\begin{tabular}[c]{@{}c@{}}State feedback\\ (poles)\end{tabular}}}                     & ${\rm P_{LK}\ \#1}$          &         [-0.93; 0.92; 0.9; 0.8]           \\ \cline{3-4} 
\multicolumn{1}{|l|}{}                              &                                                                                                                & ${\rm P_{LK}\ \#2}$          &        [-0.6$\pm$0.1i; 0.65$\pm$0.2i]            \\ \cline{3-4} 
\multicolumn{1}{|l|}{}                              &                                                                                                                & ${\rm P_{LK}\ \#3}$          &        [0.003; 0.66$\pm$0.34i; 0.4]            \\ \cline{2-4} 
\multicolumn{1}{|l|}{}                              & \multirow{3}{*}{{\begin{tabular}[c]{@{}c@{}}State feedback\\ w/ integral \\ action (poles)\end{tabular}}} & ${\rm PI_{LK}\ \#1}$          &       [-0.93; 0.92; 0.9; 0.8; 0.7]             \\ \cline{3-4} 
\multicolumn{1}{|l|}{}                              &                                                                                                                & ${\rm PI_{LK}\ \#2}$          &       [-0.6$\pm$0.1i; 0.65$\pm$0.2i; 0.7]             \\ \cline{3-4} 
\multicolumn{1}{|l|}{}                              &                                                                                                                & ${\rm PI_{LK}\ \#3}$       &          [0.002; 0.6$\pm$0.4i; 0.4; 0.7]          \\ \cline{2-4} 
\multicolumn{1}{|l|}{}                              & \multirow{3}{*}{{\begin{tabular}[c]{@{}c@{}}MPC\\ (horizon)\end{tabular}}}                              & ${\rm MPC_{LK}\ \#1}$          &          2          \\ \cline{3-4} 
\multicolumn{1}{|l|}{}                              &                                                                                                                & ${\rm MPC_{LK}\ \#2}$          &          5          \\ \cline{3-4} 
\multicolumn{1}{|l|}{}                              &                                                                                                                & ${\rm MPC_{LK}\ \#3}$          &          20          \\ \hline
\end{tabular}
\end{center}
\end{table}
}

{\small
\begin{table*}[!thpb]
\caption{ACC falsification rates (FR), with no input generation, for specifications in Eqs. \eqref{spec:ACC} and \eqref{eq:ACC_types}.}
\label{tab:ACC_PPiMpc_noinput}
\begin{center}
\begin{tabular}{|c|c|c|c|c||c|c|c|c|}
\hline
\multicolumn{1}{|c|}{\textbf{Sampling location}} & \multicolumn{4}{c|}{\textbf{Interior}} & \multicolumn{4}{c|}{\textbf{Boundary}} \\ \hline
\multicolumn{1}{|c|}{\textbf{\backslashbox{Controller}{Specification}}}&    $\varphi_{ACC}^1$  &   $\varphi_{ACC}^2$     &    $\varphi_{ACC}^3$    &$\varphi_{ACC}$  &   $\varphi_{ACC}^1$     &   $\varphi_{ACC}^2$    &   $\varphi_{ACC}^3$   &$\varphi_{ACC}$ \\ \hline\hline
${\rm P_{ACC}\ \#1}$  &0.41&0.20&0.15&0.41&0.42&0.19& 0.14&0.43\\ \cline{1-9} 
${\rm P_{ACC}\ \#2}$ &0.25	&0.14	&0.09	&0.25&0.26	&0.13	&0.09	&0.26	  \\ \cline{1-9} 
${\rm P_{ACC}\ \#3}$  &0.15	&0.11	&0.08	&0.17 &0.14	&0.11	&0.07	&0.16\\ \hline
${\rm PI_{ACC}\ \#1}$  &0.60	&0.25	&0.19	&0.60&0.63	&0.21	&0.15	&0.63	\\ \cline{1-9} 
${\rm PI_{ACC}\ \#2}$  &0.49	&0.22	&0.16	&0.49&0.41	&0.17	&0.11	&0.42	 \\ \cline{1-9} 
${\rm PI_{ACC}\ \#3}$  &0.42	&0.21	&0.14	&0.42
&0.30	&0.14	&0.10	&0.31	\\ \hline
${\rm MPC_{ACC}\ \#1}$  &0.17      & 0.13      &0.09   & 0.19   & 0.26      &  0.13     & 0.09   & 0.28  \\ \cline{1-9} 
${\rm MPC_{ACC}\ \#2}$  & 0.15     & 0.09      &0.08  &  0.15   &   0.14    &  0.09     &  0.07  & 0.14  \\ \cline{1-9} 
${\rm MPC_{ACC}\ \#3}$  & 0.15     & 0.09      &0.08  &  0.15   &   0.14    &  0.09     &  0.07  & 0.14 \\ \hline
\end{tabular}
\end{center}
\end{table*}
}

{\small
\begin{table*}[!thpb]
\caption{ACC falsification rates (FR), with dual game, for specifications in Eqs. \eqref{spec:ACC} and \eqref{eq:ACC_types}.}
\label{tab:ACC_PPiMpc_inverse}
\begin{center}
\begin{tabular}{|c|c|c|c|c||c|c|c|c|}
\hline
\multicolumn{1}{|c|}{\textbf{Sampling location}} & \multicolumn{4}{c|}{\textbf{Interior}} & \multicolumn{4}{c|}{\textbf{Boundary}} \\ \hline
\multicolumn{1}{|c|}{\textbf{\backslashbox{Controller}{Specification)}}}&    $\varphi_{ACC}^1$  &   $\varphi_{ACC}^2$     &    $\varphi_{ACC}^3$    &$\varphi_{ACC}$  &   $\varphi_{ACC}^1$     &   $\varphi_{ACC}^2$    &   $\varphi_{ACC}^3$   &$\varphi_{ACC}$ \\ \hline\hline
${\rm P_{ACC}\ \#1}$   
&1.00 	&0.88	&0.62	&1.00 	&1.00 	&0.90	&0.64	&1.00 \\ \cline{1-9} 
${\rm P_{ACC}\ \#2}$  
&1.00 	&1.00 	&0.97	&1.00 	&1.00 	&0.99	&0.98	&1.00   \\ \cline{1-9} 
${\rm P_{ACC}\ \#3}$   &0.21 	&1.00 	&0.09	&1.00 	&0.16 	&1.00	&0.08	&1.00   \\ \hline
${\rm PI_{ACC}\ \#1}$  &0.95	&0.98	&0.50	&1.00 	&0.95	&0.97	&0.55	&1.00    \\ \cline{1-9} 
${\rm PI_{ACC}\ \#2}$  &0.66	&0.99	&0.30	&1.00 	&0.54	&0.99	&0.23	&1.00    \\ \cline{1-9} 
${\rm PI_{ACC}\ \#3}$  
 	&0.55	&0.99	&0.30	&1.00 &0.39	&0.99	&0.21	&1.00  \\ \hline
${\rm MPC_{ACC}\ \#1}$ & 0.32   & 0.99 		& 0.10   &  1.00 & 0.39 & 0.99 & 0.09 & 1.00  \\ \cline{1-9} 
${\rm MPC_{ACC}\ \#2}$ & 0.20 	& 0.12 		& 0.09  &  0.20  & 0.15 & 0.09 & 0.08 & 0.15  \\ \cline{1-9} 
${\rm MPC_{ACC}\ \#3}$ & 0.20 	& 0.12 		& 0.09  &  0.20  & 0.15 & 0.09 & 0.08 & 0.15   \\ \hline

\end{tabular}
\end{center}
\end{table*}
}

{\small
\begin{table*}[!thpb]
\caption{ACC falsification rates (FR), with max braking, for specifications in Eqs. \eqref{spec:ACC} and \eqref{eq:ACC_types}.}
\label{tab:ACC_PPiMpc_maxbrake}
 \begin{center}
\begin{tabular}{|c|c|c|c|c||c|c|c|c|}
\hline
\multicolumn{1}{|c|}{\textbf{Sampling location}} & \multicolumn{4}{c|}{\textbf{Interior}} & \multicolumn{4}{c|}{\textbf{Boundary}} \\ \hline
\multicolumn{1}{|c|}{\textbf{\backslashbox{Controller}{Specification}}}&    $\varphi_{ACC}^1$  &   $\varphi_{ACC}^2$     &    $\varphi_{ACC}^3$    &$\varphi_{ACC}$  &   $\varphi_{ACC}^1$     &   $\varphi_{ACC}^2$    &   $\varphi_{ACC}^3$   &$\varphi_{ACC}$ \\ \hline\hline
${\rm P_{ACC}\ \#1}$  &1.00  & 1.00 &  1.00  & 1.00 & 1.00  & 1.00    &  1.00   & 1.00 \\ \cline{1-9} 
${\rm P_{ACC}\ \#2}$  & 1.00     & 1.00      & 1.00 & 1.00   & 1.00      & 1.00     & 1.00  &1.00    \\ \cline{1-9} 
${\rm P_{ACC}\ \#3}$  &0.25	&1.00 	&0.15	&1.00 &0.20	&1.00 	&0.13 &1.00 	 \\ \hline
${\rm PI_{ACC}\ \#1}$ &0.95&	1.00 	&0.68	&1.00 & 0.95	&1.00 	&0.72	&1.00 	  \\ \cline{1-9} 
${\rm PI_{ACC}\ \#2}$ &0.66	&1.00 	&0.43	&1.00  &0.57	&1.00 &	0.32&	1.00 	   \\ \cline{1-9} 
${\rm PI_{ACC}\ \#3}$ &0.58	&1.00 	&0.40	&1.00  &0.42	&1.00 	&0.29&	1.00 	 \\ \hline
${\rm MPC_{ACC}\ \#1}$  &0.32      & 1.00      &0.22   &  1.00  	& 0.39      &  1.00     & 0.17   	&  1.00 \\ \cline{1-9} 
${\rm MPC_{ACC}\ \#2}$  & 0.25     & 0.19      &0.15  &  0.25   	&   0.23    &  0.15     &  0.13  	& 0.23   \\ \cline{1-9} 
${\rm MPC_{ACC}\ \#3}$  & 0.25     & 0.19     &0.14  &  0.25    	&  0.28     &  0.15     &    0.13	 &0.29  \\ \hline
\end{tabular}
\end{center}
\end{table*}
}
{\small
\begin{table*}[!thpb]
\caption{ACC falsification rates (FR), with lead car converging to $\vFdes$, for specifications in Eqs. \eqref{spec:ACC} and \eqref{eq:ACC_types}.}
\label{tab:ACC_PPiMpc_accelerate}
\begin{center}
\begin{tabular}{|c|c|c|c|c||c|c|c|c|}
\hline
\multicolumn{1}{|c|}{\textbf{Sampling location}} & \multicolumn{4}{c|}{\textbf{Interior}} & \multicolumn{4}{c|}{\textbf{Boundary}} \\ \hline
\multicolumn{1}{|c|}{\textbf{\backslashbox{Controller}{Specification}}}&    $\varphi_{ACC}^1$  &   $\varphi_{ACC}^2$     &    $\varphi_{ACC}^3$    &$\varphi_{ACC}$  &   $\varphi_{ACC}^1$     &   $\varphi_{ACC}^2$    &   $\varphi_{ACC}^3$   &$\varphi_{ACC}$ \\ \hline\hline
${\rm P_{ACC}\ \#1}$   & 0.29	& 0.09	& 0.06	& 0.29 &  0.41	& 0.11	& 0.06	& 0.41	\\ \cline{1-9} 
${\rm P_{ACC}\ \#2}$  & 0.19	& 0.05	& 0.03	& 0.19 & 0.24	& 0.06	& 0.04	& 0.24	 \\ \cline{1-9} 
${\rm P_{ACC}\ \#3}$   & 0.12	& 0.04	& 0.03	& 0.12&  0.12	& 0.05	& 0.03	& 0.12 \\ \cline{1-9} 
${\rm PI_{ACC}\ \#1}$    	&0.58	&0.08	&0.05	&0.58 &0.66	&0.09	&0.05	&0.67\\ \cline{1-9} 
${\rm PI_{ACC}\ \#2}$  & 0.52	& 0.07	& 0.04	& 0.52   & 0.48	& 0.06	& 0.03	& 0.48\\ \cline{1-9} 
${\rm PI_{ACC}\ \#3}$ & 0.45	& 0.06	& 0.03	& 0.45 &  0.35	& 0.05	& 0.03	& 0.36	   \\ \cline{1-9} 
${\rm MPC_{ACC}\ \#1}$  & 0.13&  0.05& 0.03  &0.13 & 0.19  &0.06  & 0.04  & 0.20 \\ \cline{1-9} 
${\rm MPC_{ACC}\ \#2}$  &0.12 &0.04  & 0.03  & 0.12 & 0.11  &0.04   & 0.03  & 0.11  \\ \cline{1-9} 
${\rm MPC_{ACC}\ \#3}$  &0.12 & 0.04 & 0.03  & 0.12 & 0.11  & 0.04  & 0.02  & 0.11  \\ \cline{1-9} 
\end{tabular}
\end{center}
\end{table*}
}

{\small
\begin{table*}[!thpb]
\caption{LK falsification rates (FR), with input generation, for specifications in Eqs. \eqref{spec:LK} and \eqref{eq:LK_type2}.}
\label{tab:LK_PPiMpc_dual_h}
\begin{center}
\begin{tabular}{c|c|c|c|c||c|c|c|c||c|c|c|c|}
\cline{2-13}
                                                                                                 & \multicolumn{4}{c|}{\textbf{No input generation}}                               & \multicolumn{4}{c|}{\textbf{Heuristics by Eq. \eqref{eq:LK_h}}}                      & \multicolumn{4}{c|}{\textbf{Ellipsoid method + dual game}}                                        \\ \hline
\multicolumn{1}{|c|}{\textbf{Sampling location}}                                                 & \multicolumn{2}{c|}{\textbf{Interior}} & \multicolumn{2}{c|}{\textbf{Boundary}} & \multicolumn{2}{c|}{\textbf{Interior}} & \multicolumn{2}{c|}{\textbf{Boundary}} & \multicolumn{2}{c|}{\textbf{Interior}} & \multicolumn{2}{c|}{\textbf{Boundary}} \\ \hline
\multicolumn{1}{|c|}{\textbf{\backslashbox{Controller}{Specification}}} &  $\varphi_{LK}^1$      & $\varphi_{LK}$    &  $\varphi_{LK}^1$     & $\varphi_{LK}$    &  $\varphi_{LK}^1$      & $\varphi_{LK}$     &  $\varphi_{LK}^1$    & $\varphi_{LK}$     &  $\varphi_{LK}^1$     & $\varphi_{LK}$     & $\varphi_{LK}^1$     &  $\varphi_{LK}$     \\ \hline\hline
\multicolumn{1}{|c|}{${\rm P_{LK}\ \#1}$}                                            & 0.22              & 0.45            & 0.58              & 0.81             & 0.99              & 1.00             & 1.00             & 1.00             & 1.00              & 1.00             & 1.00              & 1.00             \\ \hline
\multicolumn{1}{|c|}{${\rm P_{LK}\ \#2}$}                                            & 0.00              & 0.95             & 0.00              & 0.99             & 0.00              & 1.00             & 0.00              & 1.00             & 0.00              & 1.00             & 0.00              & 1.00             \\ \hline
\multicolumn{1}{|c|}{${\rm P_{LK}\ \#3}$}                                            & 0.00              & 0.77             & 0.00              & 0.94             & 0.00              & 1.00             & 0.00              & 1.00             & 0.00              & 1.00             & 0.00              & 1.00             \\ \hline
\multicolumn{1}{|c|}{\textbf{${\rm PI_{LK}\ \#1}$}}                                  & 0.00              & 0.33             & 0.00              & 0.71             & 1.00              & 1.00             & 1.00              & 1.00             & 0.38              & 0.65             & 0.63              & 0.91             \\ \hline
\multicolumn{1}{|c|}{\textbf{${\rm PI_{LK}\ \#2}$}}                                  & 0.00              & 0.99             & 0.03              & 1.00             & 0.14             & 1.00             & 0.26              & 1.00             & 0.08              & 1.00             & 0.15              & 1.00             \\ \hline
\multicolumn{1}{|c|}{\textbf{${\rm PI_{LK}\ \#3}$}}                                  & 0.00              & 0.95             & 0.00              & 0.99             & 0.00              & 1.00             & 0.01              & 1.00             & 0.05              & 0.99             & 0.10              & 1.00             \\ \hline
\multicolumn{1}{|c|}{\textbf{${\rm MPC_{LK}\ \#1}$}}                                 & 0.00              & 0.00             & 0.00              & 0.00             & 1.00              & 1.00             & 1.00              & 1.00             & 1.00              & 1.00             & 1.00              & 1.00             \\ \hline
\multicolumn{1}{|c|}{\textbf{${\rm MPC_{LK}\ \#2}$}}                                 & 0.00              & 0.00             & 0.00              & 0.00             & 0.04              & 0.04             & 0.09              & 0.09             & 0.04              & 0.04             & 0.08              & 0.08 \\ \hline
\multicolumn{1}{|c|}{\textbf{${\rm MPC_{LK}\ \#3}$}}                                 & 0.00              & 0.00             & 0.00              & 0.00             & 0.01              & 0.11            & 0.01              & 0.44             & 0.002              & 0.11             & 0.01              & 0.44             \\ \hline
\end{tabular}
\end{center}
\end{table*}
}

{\small
\begin{table*}[!thpb]
\caption{ACC falsification rates (FR): \comma{}, for specifications in Eqs. \eqref{spec:ACC} and \eqref{eq:ACC_types}.}
\label{tab:ACC_comma}
\begin{center}
\begin{tabular}{|c|c|c|c|c||c|c|c|c|}
\hline
\textbf{Sampling location} & \multicolumn{4}{c|}{\textbf{Interior}} & \multicolumn{4}{c|}{\textbf{Boundary}} \\ \hline
\textbf{\backslashbox{Input generation}{Specification}}&    $\varphi_{ACC}^1$  &  $\varphi_{ACC}^2$   &   $\varphi_{ACC}^3$  &   $\varphi_{ACC}$ &  $\varphi_{ACC}^1$    &   $\varphi_{ACC}^2$   &    $\varphi_{ACC}^3$  &  $\varphi_{ACC}$  \\ \hline\hline
\textbf{No input generation}  & 0.15     &  0.00     &  0.00  &0.15   & 0.42      &  0.04     &  0.00 & 0.44     \\ \hline
\textbf{Heuristic (max brake)} &   0.15    &  0.29      & 0.00  &0.29 & 0.39    &  0.43      & 0.00 &0.48 \\ \hline
\textbf{Ellipsoid method + dual game} & 0.15      &  0.17     &  0.00 &0.29  & 0.36    & 0.19     & 0.00 &0.48 \\ \hline
\textbf{aLead = K(vLead -v\_des)} & 0.43     &  0.00     &  0.00 &0.43  & 0.56    & 0.01     & 0.00&0.56  \\ \hline
\end{tabular}
\end{center}
\end{table*}
}

{\small
\begin{table*}[!thpb]
\caption{LK falsification rates (FR): \comma{} \& \commastar{} (without modification), for specifications in Eqs. \eqref{spec:LK} and \eqref{eq:LK_type2}.}
\label{tab:LK_comma_comma_star}
\begin{center}
\begin{tabular}{c|c|c|c|c||c|c|c|c|}
\cline{2-9}
                                                            & \multicolumn{4}{c|}{\textbf{\comma{}}}                                                 & \multicolumn{4}{c|}{\textbf{\commastar{} (without modification)}}                                                \\ \hline
\multicolumn{1}{|c|}{\textbf{Sampling location}}            & \multicolumn{2}{c|}{\textbf{Interior}} & \multicolumn{2}{c|}{\textbf{Boundary}} & \multicolumn{2}{c|}{\textbf{Interior}} & \multicolumn{2}{c|}{\textbf{Boundary}} \\ \hline
\multicolumn{1}{|c|}{\textbf{\backslashbox{Input   generation}{Specifcation}}}                             & $\varphi_{LK}^1$     &  $\varphi_{LK}$     &  $\varphi_{LK}^1$     &  $\varphi_{LK}$     & $\varphi_{LK}^1$      &  $\varphi_{LK}$     & $\varphi_{LK}^1$     &  $\varphi_{LK}$     \\ \hline\hline
\multicolumn{1}{|c|}{\textbf{No input generation}}          & 0.002              & 0.868             & 0.034              & 0.968             & 0                  & 1                 & 0.014              & 1                 \\ \hline
\multicolumn{1}{|c|}{\textbf{Heuristic by Eq. \eqref{eq:LK_h}}}                    & 0.000              & 0.880             & 0.014              & 0.982             & 0.272              & 0.956             & 0.264              & 0.996             \\ \hline
\multicolumn{1}{|c|}{\textbf{Ellipsoid method + dual game}} & 0.006              & 0.952             & 0.038              & 0.994             & 1                  & 1                 & 1                  & 1                 \\ \hline
\end{tabular}
\end{center}
\end{table*}
}

\subsection{\comma{}}

Our framework is flexible enough to evaluate any type of controller as long as their inputs and outputs match the inputs and outputs of the system models used. We can also directly use the source code of a controller after developing a proper interface. In this section, we demonstrate our framework on an open source real-world autonomous driving package developed by \comma{}, a start-up working on self-driving car technologies (see \url{https://comma.ai/}). However, since we are just using a simplified model for the vehicle dynamics, the interface might not accurately reflect the performance of the software on an actual car. This can be improved by improving the models and interfaces, but the goal in this section is to simply show the applicability of the framework on realistic control software rather than accurately mimicking the performance.

We describe how we interfaced the \comma{} code (commit \texttt{5524dc8}\footnote{After we settled on a version of \comma{} to use for this project, newer versions of the \comma{} code have changed the lane-keeping module from using a PI to an MPC-based controller. Testing this new controller within our framework is the subject of future work.} at \url{https://github.com/commaai}) with our ACC and LK framework. The \comma{} code is written in Python. We call the Python code directly from within Matlab by developing appropriate wrappers for input/output matching as described next. 

The ACC module of \comma{} outputs two values: gas and brake commands, each normalized to $[0, 1]$. We scale these gas and brake commands by the physical gas and brake limits of average mid-sized sedans, $\Fwmaxphys$ and $\Fwminphys$, respectively. We then clip the scaled gas and brake commands to the comfort bounds $[0, F_w^\textrm{max}]$ and $[F_w^\textrm{max}, 0]$, respectively. The sum of the scaled gas and brake commands is used as the control input to our system.

The LK module of \comma{} requires extra interfacing with our simulations. 
Firstly, \comma{} outputs a control between $u\in [-1, 1]$, and we assume that these bounds map linearly onto the range of steering angles $[\theta_s^{\rm min}, \theta_s^{\rm max}]$. Secondly, \comma{} takes as input the road profile, at $dR_c$ discretization, for the upcoming 50 m, which is measured along the tangent line of the current car configuration. To provide this at time step $T$,

\begin{enumerate}
\item We compute a sequence of future road curvature disturbances $r_d^{1:n}(T) \doteq \{ d_t\}_{t=1}^n(T)$, where $n \ge 50$ using one of the input generation methods. To be consistent with prior road profiles, we fix $r_d^{1:n-1}(T) = r_d^{2:n}(T-1)$ and only compute $r_d^{n}(T)$ from scratch. In principle, we can compute $r_d^{1:n}(T)$ entirely from scratch; this setting can be interpreted as driving with \comma{} vision sensor failure/noise, leading to inconsistent roads from prior time steps. However, by ensuring that we provide consistent roads, we give \comma{} the advantage here by assuming that the vision data is exact.
\item Assuming that the $r_d^{1:n}(T)$ trajectory was obtained from measurements taken at a rate of $dT_c$ of the vehicle traveling at $v_N$ m/s on the center line, we can estimate the center line in road-fixed coordinates, $R_d^{1:n}(T)$, by approximating that the road traces out arcs of angle $r_d^i dT_c$ at every time step $i$. That is, 
\begin{equation*}
  R_d^i(T) = R_d^{i-1}(T) + \frac{v_N}{r_d^i(T)}  \begin{bmatrix}
    \cos\Big(\frac{r_d^i(T)}{v_N} - \frac{\pi}{2}\Big) \\
    \sin\Big(\frac{r_d^i(T)}{v_N} - \frac{\pi}{2}\Big) + 1
  \end{bmatrix}.
\end{equation*}
\item Since $R_d$ is relative to a road-fixed coordinate system, we rotate and translate $R_d$ into the car frame by rotating each waypoint by $-\Delta \Psi$ and translating by $-y$ (denoted $R_d'(T)$).
\item We evaluate $R_d'(T)$ values at a discretization of 1 m along the tangent line using linear interpolation.
\end{enumerate}

Additionally, we modified one of the vehicle model equations in the original \comma{} code (see Figure \ref{fig:bug_fix}). We will refer to the original \comma{} code as \commastar{}, and to our modified code as \comma{}, which performs better with our interface. 

\begin{figure}[t] 
	\centering 
	\includegraphics[width=0.95\linewidth]{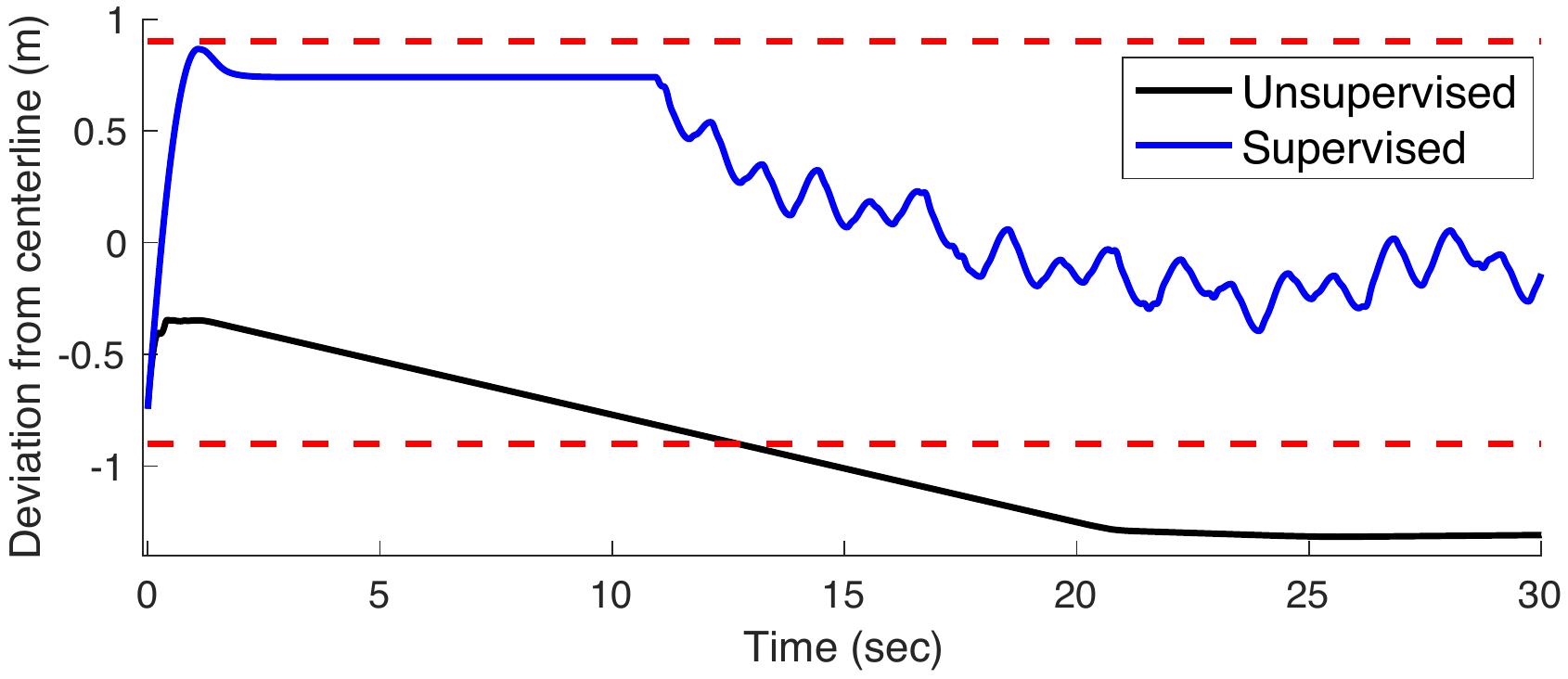}
	\caption{Un/supervised LK trajectories using \commastar{}}\label{fig:comma_star} 
\end{figure}

Figures \ref{fig:comma} and \ref{fig:comma_star} show a trajectory generated by \comma{} and \commastar{}, respectively, that leaves the lane boundaries, overlaid by the trajectory generated by \comma{} and \commastar{} when they are used as the legacy controller when the invariant set based supervisor is active.

We see that for ACC, \comma{} manages to stay out of crashes for all initial conditions and input generation method; however, it violates time and distance headways many times, as Table \ref{tab:ACC_comma} suggests. These violations are undesirable since the passengers might feel uncomfortable when \comma{} follows the lead car too closely. Furthermore, the \comma{} code itself sets a soft constraint for the desired distance headway being greater than 4m, which is frequently violated.

The LK statistics in the left half of Tables \ref{tab:LK_comma_comma_star} indicate that while \comma{} stays within the lane boundaries for the most part when starting from nonzero initial conditions, in the process of stabilization, it tends to violate comfort bounds. Falsification is more likely starting from initial conditions in the boundary than in the interior. \commastar{} (see the right half of Table \ref{tab:LK_comma_comma_star}) drives slightly better on straight roads but performs much worse with input generation compared to \comma{}, which is consistent with the decent performance of \commastar{} on simple roads. The ellipsoid + dual game method of input generation seems to falsify \comma{} and \commastar{} more than the heuristic method; in fact, a straight road tends to falsify \comma{} and \commastar{} more than the heuristic method. This happens because the heuristic method tends to smooth out the natural overshoot that \comma{} and \commastar{} exhibit in their responses.

\begin{figure}[t] 
	\centering 
	\includegraphics[width=0.95\linewidth]{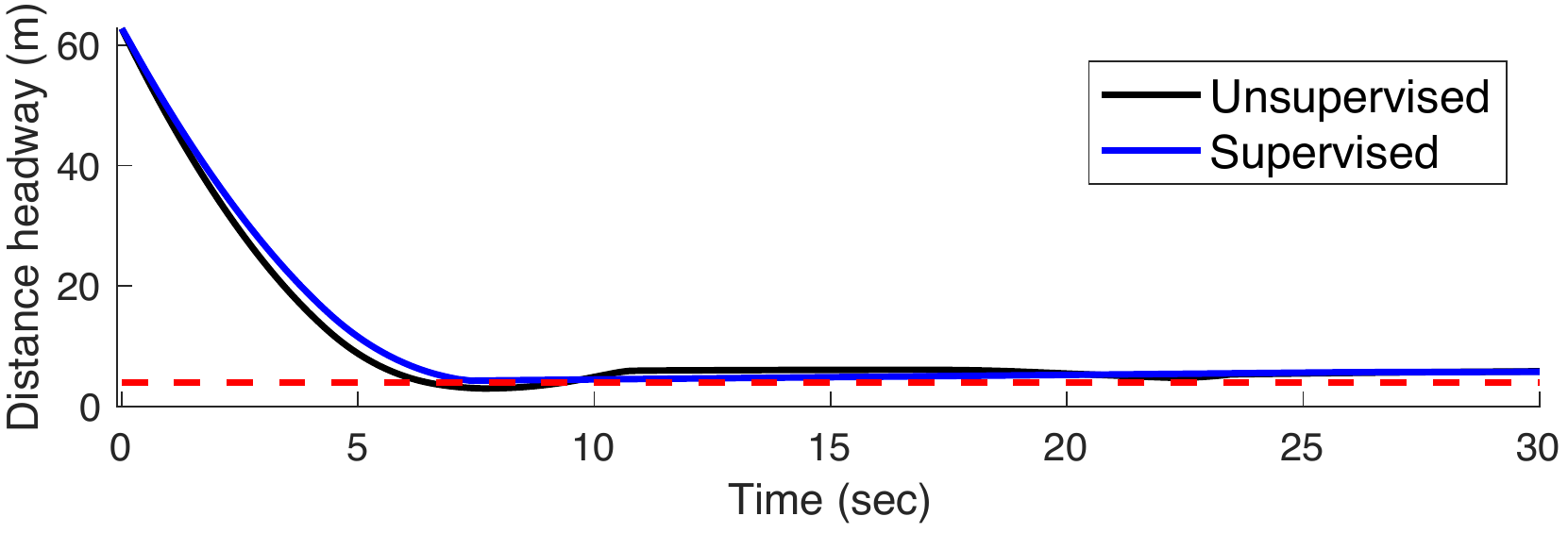}
	\caption{Un/supervised ACC trajectories using \comma{}}\label{fig:comma} 
\end{figure}

\begin{figure}[t] 
	\centering 
	\includegraphics[width=0.95\linewidth]{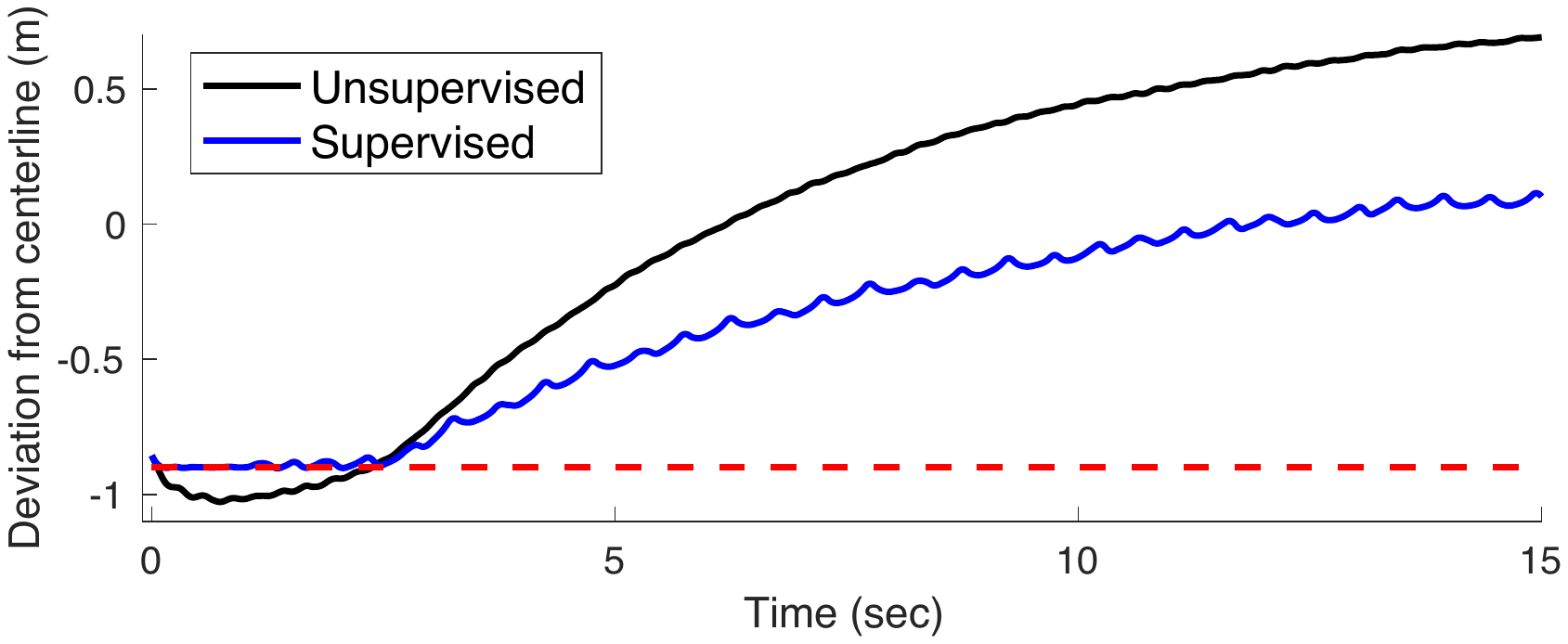}
	\caption{Un/supervised LK trajectories using \comma{}}\label{fig:comma} 
\end{figure}

\subsection{\staliro{} Results}
For comparison and benchmarking purposes, we use \staliro{}, a falsification tool that is proposed in \cite{annpureddy2011s}, to find falsifying initial conditions and disturbance trajectories. Although \staliro{} and our approach are somewhat complementary, we try to demonstrate some of the differences. First, note that \staliro{} does not provide any information about whether a falsifying initial condition disturbance pair is ``interesting", so it is not known if the violation is due to poor performance of the controller or it is unavoidable.  To demonstrate to what extent \staliro{} can find ``interesting" falsifying trajectories, we use the ACC example. We restrict the initial conditions that \staliro{} can choose by upper-bounding the headway $h\leq 200m$. Let, $X_0 :=  [\vFmin,\vFmax]\times[\hmin, 200]\times[\vLmin,\vLmax]$.
Then, the specification used in \staliro{} for falsification is 
\begin{align}\label{spec:staliro}
\nonumber\Big( \big((v(0),h(0), v_{L}(0))&\in S\cap X_0\big) \land \\ 
&\big(\forall t: 
\vL(t) \geq \vLmin \big)\Big) \rightarrow  \varphi^1_{ACC} \wedge \varphi^2_{ACC}.
\end{align}  
In addition, we impose bounds on the external inputs, i.e., $\aL(t) \in [\aLmin,\aLmax]$ for all $t$; and the domain of the dynamics is accounted for by the simulation model. Note that by the assumptions on the initial conditions and $\vL$ in \eqref{spec:staliro}, we avoid some of the trivially unsafe falsifications.
 To falsify $P$ and $PI$ controllers and \comma{}, we limit the number of samples \staliro{} can try to find a falsifying trajectory to $100$; this acts as a timeout condition. We then run \staliro{} for $100$ times with the default option of simulated annealing, a random search method. Table \ref{tab:acc_staliro} summarizes the results, showing not all falsifying trajectories found by \staliro{} are ``interesting''. Furthermore, \staliro{} sometimes fails to falsify $P$ and $PI$ controllers before our timeout condition, whereas Table~\ref{tab:ACC_PPiMpc_inverse} shows that the dual game approach always finds inputs that lead to falsification.

We also provide \staliro{} with initial conditions sampled from the invariant set boundary, therefore forcing it to find ``interesting" initial conditions. These results are reported in the last column of Table~\ref{tab:acc_staliro}. Although \staliro{} finds fewer falsifying trajectories for some controllers in this case, all of the falsifying trajectories found are ``interesting" by definition.
Comparing the last column in Table~\ref{tab:acc_staliro} with that in Tables~\ref{tab:ACC_PPiMpc_inverse} and \ref{tab:ACC_comma}, we see that our input generation does better for most controllers except for \comma{}, for which \staliro{} has a slightly higher falsification rate.  

Finally, we give \staliro{} $111$ initial conditions from the winning set of the dual game. \staliro{} takes 43secs to falsify all the points whereas dual game takes 170secs. This is partly due to the fact that the inputs selected by dual game input generation are arbitrary (within the winning inputs) but not necessarily aggressive, which can be mitigated by including an objective function in input generation phase. It is also worth mentioning that for these initial conditions, our approach is guaranteed to find a falsifying trajectory however \staliro{} does not have such a guarantee due to its random nature.

{\small
\begin{table}[!thpb]
\caption{ACC falsification with  \staliro{}. The Falsified column shows the fraction of times \staliro{} finds a falsifying trajectory/initial condition pair for the specification \eqref{spec:staliro} before timing out. The invariant set column shows the fraction of times a falsifying pair with an ``interesting'' initial condition is found among all runs. The last column shows the falsification rate when \staliro{} is given initial conditions sampled from the boundary of the invariant set.}
\label{tab:acc_staliro}
\begin{center}
\begin{tabular}{|c|c|c|c|}
\cline{1-4}
\multirow{ 2}{*}{\textbf{Controller}} &  \multicolumn{2}{c|}{\textbf{\staliro{}}} & \textbf{\staliro{} with IC's}    \\
\cline{2-3}
 & \textbf{Falsified}      &         \textbf{In invariant set} & \textbf{on Boundary} \\\hline \hline
${\rm P_{ACC}\ \#1}$ &    0.95      & 0.89             & 1.00 \\ \cline{1-4} 
${\rm P_{ACC}\ \#2}$ &    0.94      &     0.90          &0.72 \\ \cline{1-4} 
${\rm P_{ACC}\ \#3}$ &    0.93      &     0.88         & 0.69 \\ \hline
${\rm PI_{ACC}\ \#1}$ &    0.97      &   0.91         & 0.96\\ \cline{1-4} 
${\rm PI_{ACC}\ \#2}$ &    0.97      &   0.93        & 0.87 \\ \cline{1-4} 
${\rm PI_{ACC}\ \#3}$ &    0.96      &    0.91       &  0.89\\ \hline
${\comma}$ &   0.68       &   0.51          &0.58  \\\hline
\end{tabular}
\end{center}
\end{table}
}

\section{Conclusions and Discussion}
\label{sec:conclusions}
This paper proposed a simple idea on how to use controlled invariant sets and solutions from a dual game to generate interesting corner cases that can be used for falsification of safety specifications.
We illustrated the effectiveness of this idea with an extensive case study on two autonomous driving functions, namely adaptive cruise control and lane keeping, with various types of controllers, including an open source autonomous driving package \comma{}. Our simulations show that we can identify corner cases with synthesis techniques and also supervise existing controllers to avoid failure in such corner cases.

The proposed approach should not be considered as an alternative to falsification techniques, as it is limited to safety specifications and to cases where approximating the maximal invariant set is possible. Therefore, it requires some knowledge of the system dynamics although it is agnostic to the controller. Whereas, advanced falsification engines \cite{annpureddy2011s} can handle rich specifications given in signal temporal logic, and even black-box system models. On the other hand, we believe our approach can be used to seed falsification engines by applying it to the safety part of a specification, a direction for future research.

\section*{Acknowledgment}
Toyota Research Institute ("TRI") provided funds to assist the authors with their research but this article solely reflects the opinions and conclusions of its authors and not TRI or any other Toyota entity.

\bibliographystyle{abbrv}
\bibliography{ref}

\begin{thebibliography}{10}

\bibitem{isoACCstandard}
{\em Intelligent Transport Systems-Adaptive Cruise Control Systems- Performance
  Requirements and Test Procedures}, ISO Standard 15622:2010 (E), 2010.

\bibitem{annpureddy2011s}
Y.~Annpureddy, C.~Liu, G.~E. Fainekos, and S.~Sankaranarayanan.
\newblock S-taliro: A tool for temporal logic falsification for hybrid systems.
\newblock In {\em TACAS}, volume 6605, pages 254--257. Springer, 2011.

\bibitem{bak2011sandboxing}
S.~Bak, K.~Manamcheri, S.~Mitra, and M.~Caccamo.
\newblock Sandboxing controllers for cyber-physical systems.
\newblock In {\em Proceedings of the 2011 IEEE/ACM Second International
  Conference on Cyber-Physical Systems}, pages 3--12. IEEE Computer Society,
  2011.

\bibitem{blanchini1999survey}
F.~Blanchini.
\newblock Survey paper: Set invariance in control.
\newblock {\em Automatica}, 35(11):1747--1767, 1999.

\bibitem{boender1991shakeandbake}
C.~G.~E. Boender, R.~J. Caron, J.~F. McDonald, A.~H. G.~R. Kan, H.~E. Romeijn,
  R.~L. Smith, J.~Telgen, and A.~C.~F. Vorst.
\newblock Shake-and-bake algorithms for generating uniform points on the
  boundary of bounded polyhedra.
\newblock {\em Operations Research}, 39(6):945--954, 1991.

\bibitem{DeSantis:2004et}
E.~De~Santis, M.~D. Di~Benedetto, and L.~Berardi.
\newblock {Computation of Maximal Safe Sets for Switching Systems}.
\newblock {\em IEEE Trans. Autom. Control}, 49(2):184--195, 2004.

\bibitem{fainekos2012verification}
G.~E. Fainekos, S.~Sankaranarayanan, K.~Ueda, and H.~Yazarel.
\newblock Verification of automotive control applications using s-taliro.
\newblock In {\em American Control Conference (ACC), 2012}, pages 3567--3572.
  IEEE, 2012.

\bibitem{ghosh2013nearly}
M.~Ghosh, H.-Y.~C. Yeh, S.~Thomas, and N.~M. Amato.
\newblock Nearly uniform sampling on surfaces with applications to motion
  planning.
\newblock 2013.

\bibitem{herceg2013multi}
M.~Herceg, M.~Kvasnica, C.~N. Jones, and M.~Morari.
\newblock Multi-parametric toolbox 3.0.
\newblock In {\em Control Conference (ECC), 2013 European}, pages 502--510.
  IEEE, 2013.

\bibitem{john2014extremum}
F.~John.
\newblock Extremum problems with inequalities as subsidiary conditions.
\newblock In {\em Traces and emergence of nonlinear programming}, pages
  197--215. Springer, 2014.

\bibitem{Leydold98asweep-plane}
J.~Leydold and W.~Hörmann.
\newblock A sweep-plane algorithm for generating random tuples in simple
  polytopes.
\newblock {\em MATHEMATICS OF COMPUTATION}, 67(224):1617--1635, 1998.

\bibitem{pn_thesis2017}
P.~Nilsson.
\newblock {\em Correct-by-Construction Control Synthesis for High-Dimensional
  Systems}.
\newblock PhD thesis, University of Michigan, 2017.

\bibitem{nilsson2016correct}
P.~Nilsson, O.~Hussien, A.~Balkan, Y.~Chen, A.~D. Ames, J.~W. Grizzle, N.~Ozay,
  H.~Peng, and P.~Tabuada.
\newblock Correct-by-construction adaptive cruise control: Two approaches.
\newblock {\em IEEE Trans. on Control Systems Technology}, 24(4):1294--1307,
  2016.

\bibitem{ozay2017guest}
N.~Ozay and P.~Tabuada.
\newblock Guest editorial: special issue on formal methods in control.
\newblock {\em Discrete Event Dynamic Systems}, 27(2):205--208, 2017.

\bibitem{prajna2004safety}
S.~Prajna and A.~Jadbabaie.
\newblock Safety verification of hybrid systems using barrier certificates.
\newblock In {\em HSCC}, volume 2993, pages 477--492. Springer, 2004.

\bibitem{rimon1992efficient}
E.~Rimon and S.~P. Boyd.
\newblock Efficient distance computation using best ellipsoid fit.
\newblock In {\em Intelligent Control, 1992., Proceedings of the 1992 IEEE
  International Symposium on}, pages 360--365. IEEE, 1992.

\bibitem{roy2011pessoa}
P.~Roy, P.~Tabuada, and R.~Majumdar.
\newblock Pessoa 2.0: a controller synthesis tool for cyber-physical systems.
\newblock In {\em Proceedings of the 14th international conference on Hybrid
  systems: computation and control}, pages 315--316. ACM, 2011.

\bibitem{rungger2017computing}
M.~Rungger and P.~Tabuada.
\newblock Computing robust controlled invariant sets of linear systems.
\newblock {\em IEEE Transactions on Automatic Control}, 2017.

\bibitem{sankaranarayanan2012falsification}
S.~Sankaranarayanan and G.~Fainekos.
\newblock Falsification of temporal properties of hybrid systems using the
  cross-entropy method.
\newblock In {\em Proceedings of the 15th ACM international conference on
  Hybrid Systems: Computation and Control}, pages 125--134. ACM, 2012.

\bibitem{seto1998dynamic}
D.~Seto, B.~H. Krogh, L.~Sha, and A.~Chutinan.
\newblock Dynamic control system upgrade using the simplex architecture.
\newblock {\em IEEE Control Systems}, 18(4):72--80, 1998.

\bibitem{smith1984pointsuniform}
R.~L. Smith.
\newblock Efficient monte carlo procedures for generating points uniformly
  distributed over bounded regions.
\newblock {\em Operations Research}, 32(6):1296--1308, 1984.

\bibitem{smith2016interdependence}
S.~W. Smith, P.~Nilsson, and N.~Ozay.
\newblock Interdependence quantification for compositional control synthesis
  with an application in vehicle safety systems.
\newblock In {\em Decision and Control (CDC), 2016 IEEE 55th Conference on},
  pages 5700--5707. IEEE, 2016.

\end{thebibliography}




\end{document}